\documentclass[12pt]{article}
\usepackage{geometry,graphicx,natbib,amsfonts,verbatim,mathrsfs,amssymb} 

\usepackage{longtable,wrapfig,fullpage,setspace}
\usepackage{hyperref}
\usepackage{color}
\usepackage{authblk}
\usepackage{subcaption}
\hypersetup{colorlinks,linkcolor=blue,citecolor=blue}

\newcommand{\be}{\begin{equation}}  % 
\newcommand{\ee}{\end{equation}}  % 
\newcommand{\ohmm}{\Omega^{-1}~\mbox{m}^{-1}}
\newcommand{\jmol}{\mbox{J~mol}^{-1}} 

\newcommand{\tidalq}{\mathcal{Q}}
\title{Tidal heating of Earth-like exoplanets around M stars: Thermal, magnetic, and orbital evolutions}
\author[1,2]{Driscoll, P. E.\thanks{ped13@uw.edu, (206) 543-0777}}
\author[1,2]{Barnes, R.}
\affil[1]{Astronomy Department, University of Washington, Seattle, WA}
\affil[2]{NASA Astrobiology Institute, Virtual Planet Laboratory Lead Team}
\date{\small Compiled \today \\
Accepted in \\  Astrobiology, Volume 15, Number 9, 2015 \\  DOI: 10.1089/ast.2015.1325 \\
 Note: Small deviations exist between this version \\ and the published version. }

\begin{document}

\maketitle

\abstract{
The internal thermal and magnetic evolution of rocky exoplanets is critical to their habitability.  We focus on the thermal-orbital evolution of Earth-mass planets around low mass M stars whose radiative habitable zone overlaps with the ``tidal zone", where tidal dissipation is expected to be a significant heat source in the interior.  We develop a thermal-orbital evolution model calibrated to Earth that couples tidal dissipation, with a temperature-dependent Maxwell rheology, to orbital circularization and migration.  We illustrate thermal-orbital steady states where surface heat flow is balanced by tidal dissipation and cooling can be stalled for billions of years until circularization occurs.  Orbital energy dissipated as tidal heat in the interior drives both inward migration and circularization, with a circularization time that is inversely proportional to the dissipation rate.  We identify a peak in the internal dissipation rate as the mantle passes through a visco-elastic state at mantle temperatures near 1800 K.  Planets orbiting a 0.1 solar-mass star within $0.07$ AU circularize before 10 Gyr, independent of initial eccentricity.  Once circular, these planets cool monotonically and maintain dynamos similar to Earth. Planets forced into eccentric orbits can experience a super-cooling of the core and rapid core solidification, inhibiting dynamo action for planets in the habitable zone.  We find that tidal heating is insignificant in the habitable zone around $0.45$ (or larger) solar mass stars because tidal dissipation is a stronger function of orbital distance than stellar mass, and the habitable zone is further from larger stars.  Suppression of the planetary magnetic field exposes the atmosphere to stellar wind erosion and the surface to harmful radiation. In addition to weak magnetic fields, massive melt eruption rates and prolonged magma oceans may render eccentric planets in the habitable zone of low mass stars inhospitable for life.
}


\section{Introduction}
Gravitational tides are common in the Solar System, from the Moon, responsible for driving the principle diurnal tides in Earth's oceans and atmosphere, to Io, the most volcanically active body in the Solar System.
Tidal dissipation as a heat source in the solid Earth is weak at present and often neglected from thermal history calculations of its interior.  However, rocky exoplanets with eccentric orbits close to their star are expected to experience significant tides \citep{dole1964,rasio1996,jackson2009,barnes2010} that likely influence their thermal, orbital, and even atmospheric evolution \citep{barnes2013,luger2015}.  Recent progresses in modeling tidal dissipation in a visco-elastic mantle \citep{behounkova2010,behounkova2011,henning2009,henning2014} have advocated using Maxwell-type temperature-pressure dependent rheology and emphasized the limited applicability of a constant tidal quality factor ``$\mathcal{Q}$'' model.  These more complicated dissipation models are necessary to better characterize the tidal and orbital states of rocky exoplanets over a range of internal temperatures.

The search for habitable Earth--like exoplanets commonly targets planets in orbit around low mass M--type stars to maximize the number of small mass planets found \citep{mayor2014,seager2013}.  Targeting low mass stars is beneficial for at least three reasons: (1) the habitable zone around M stars is much closer to the star \citep{kopparapu2013}, making an Earth--mass planet in the habitable zone an easier target for both transit and radial velocity detection, (2) low mass M stars are more abundant in the nearby solar neighborhood, and (3) M stars have longer main sequence times.  

On the other hand, there are several reasons why targeting M stars may be risky.  {  M stars are intrinsically faint making most observations low signal to noise, and their flux peaks at wavelengths close to those absorbed by Earth's atmosphere.}
M stars are more active, especially early on, which may induce massive amounts of atmospheric loss \citep{luger2015} and biologically hazardous levels of radiation at the surface.  Earth--mass planets in the habitable zones of M stars likely experience larger gravitational tides associated with star-planet and planet-planet interactions, especially considering that most exoplanet systems are dynamically full \citep{barnes2004,barnes2006,vanlaerhoven2014}.  {  However, the implications of these tides on the thermal evolution of the interior have not yet been explored.}

The thermal history of a planet is critical to its habitability.  Mantle temperature determines the rates of melting, degassing, and tectonics, while the thermal state of the core is critical to the maintenance of a planetary magnetic field that shields the surface from high energy radiation.
    The thermal evolution of the Earth and terrestrial planets involves solving the time evolution of mantle and core temperature through a balance of heat sources and sinks.  The thermal history of the Earth, although better constrained than any other planet, is still subject to significant uncertainties.  However, avoiding both the thermal catastrophe in the mantle \citep{korenaga2006} and the new core paradox \citep{olson2013} add significant constraints that predict a monotonic cooling of the mantle and an active geodynamo over the history of the planet \citep{driscoll2014}.      Previous models of the thermal evolution of rocky exoplanets \citep[e.g.][]{gaidos2010,driscoll2011a,tachinami2011,vansummeren2013,zuluaga2013} have focused on the influence of planet size on thermal evolution, but neglected tidal dissipation as an internal heat source, and therefore likely over estimate magnetic field strength and lifetime in the HZ around low mass stars.  We improve the thermal and magnetic evolution model used in \citet{driscoll2014} by adding tidal heating as a internal heat source, and couple this to orbital evolution.

In this paper we focus on the influence of tidal dissipation in the solid mantle of Earth--like exoplanets in the habitable zone around M stars.  Tidal dissipation deposits heat in the planetary interior while simultaneously extracting energy from the orbit, which can lead to circularization and migration.  We couple the thermal and orbital evolution equations into a single model to identify the conditions and timescales for Earth--like geophysical and magnetic evolution.  Section \S \ref{section_model} describes the thermal-orbital model equations.  Steady state behavior is discussed in \S \ref{section_ss} to build intuition about the thermal-orbital coupling.  Results with evolving orbits are presented in \S \ref{section_evolvingorbits} and with fixed orbits, mimicking forcing by companion planets, in \S \ref{section_fixedorbits}.  
{  The possibility of an internally driven runaway greenhouse is addressed in \S \ref{section_runaway}.  The influence of tides on the inner edge of the habitable zone over a range of stellar masses is explored in \S \ref{section_stellarmass}.}
A summary and discussion are in \S \ref{section_discussion}.

%%%%%%  MODEL DESCRIPTION   %%%%%%%%%%%%%%%%%%%%%%%
\section{Model Description} \label{section_model}
In this section we describe the details of the thermal-orbital evolution model.  \S \ref{section_tidal_model} describes the tidal dissipation model, \S \ref{section_thermal_model} describes the thermal evolution of the coupled mantle-core interior, and \S \ref{section_orbital_model} describes the orbital evolution as a function of dissipation efficiency.  A note on terminology: here $Q$'s refer to heat flows (units of [W]), $q$'s refer to heat fluxes (units of [W m$^{-2}$]), and the script symbol $\tidalq$ refers to the tidal quality factor (dimensionless), also known as the specific dissipation function.

%%%%%%  TIDAL MODEL   %%%%%%%%%%%%%%%%%%%%%%%
\subsection{Tidal Dissipation Model}  \label{section_tidal_model}
The gravitational perturbation experienced by a secondary body (the planet) in orbit about a primary body (the star) is approximated by the lowest order term in the potential expansion, which is the semidiurnal tide of degree 2 \citep{kaula1964}.
The power dissipated by tidal strain associated with this term in the secondary with synchronous rotation is \citep{segatz1988},
\be  Q_{tidal}=-\frac{21}{2} \mbox{Im}(k_2) \frac{G M_*^2 R_p^5 \omega e^2}{a^6}  \label{q_tidal} \ee
where $G$ is the gravitational constant, $M_*$ is stellar mass, $R_p$ is planet radius, $\omega$ is orbital frequency, $e$ is orbital eccentricity, $a$ is orbital semi-major axis, and $\mbox{Im}(k_2)$ is the imaginary part of the complex second order love number $k_2$.
If planetary rotation is synchronous then the tidal frequency is equal to the mean motion $\omega=n=\sqrt{GM_*/a^3}$, and the tidal power reduces to
\be  Q_{tidal}= - \frac{21}{2}  \mbox{Im}(k_2)G^{3/2} M_*^{5/2} R_p^5 \frac{e^2}{a^{15/2}}  ~. \label{q_tidal} \ee
This expression for tidal dissipation is the product of three physical components: (1) tidal efficiency ($-$Im$(k_2)$), (2) star-planet size ($M_*^{5/2}R_p^5$), and (3) orbit ($e^2/a^{15/2}$).

\begin{figure} \begin{center}
\includegraphics[width=0.8\linewidth]{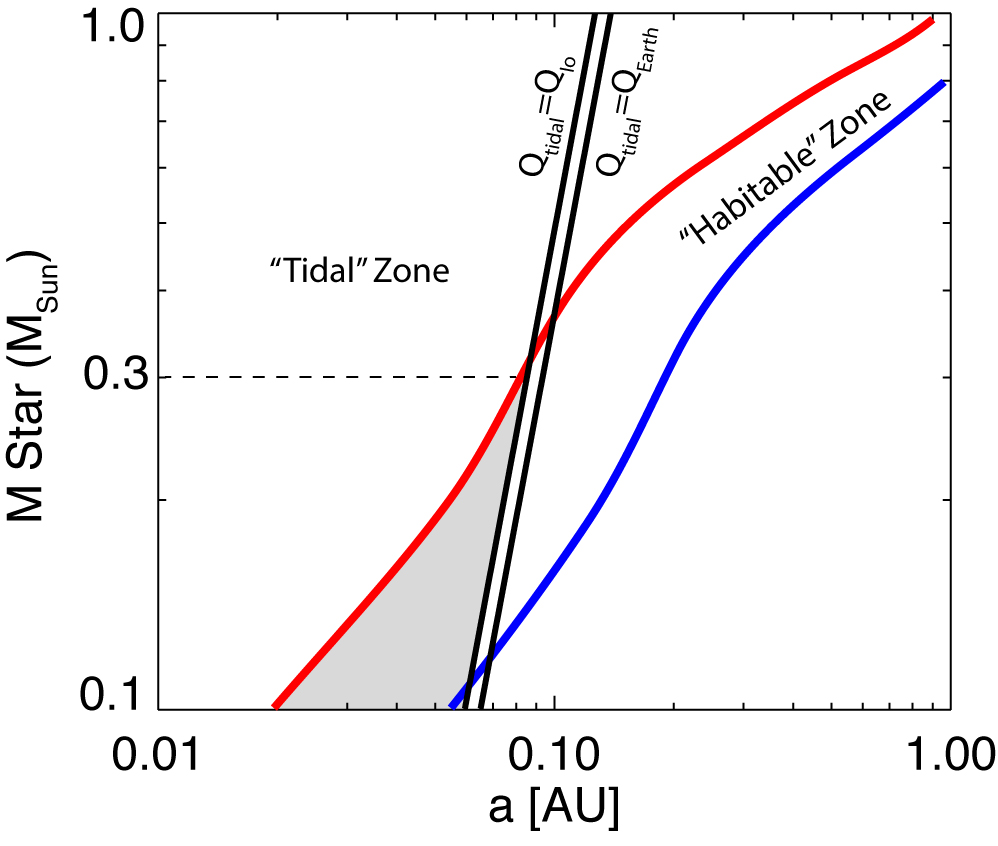}
\caption{Comparison of the radiative "habitable" zone to the "tidal" zone.  The radiative ``habitable" zone is from \citet{kopparapu2013}.  Inside the ``tidal" zone heat released by tidal dissipation is likely to dominate the internal heat budget of the planet.  The tidal zone is delineated by distances from the star where an Earth-mass planet would receive an amount of heat via tidal dissipation equal to either the surface heat flow of Io ($Q_{surf}\approx80$ TW, left curve) or Earth ($Q_{surf}=40$ TW, right curve).  Tidal heat flow is calculated by (\ref{q_tidal}) assuming $e=0.1$ and $-\mbox{Im}(k_2)=3\times10^{-3}$ ($k_2=0.3$, $\mathcal{Q}=100$).  The gray shaded region denotes the zone where the planet is predicted to be radiatively ``habitable" but tidally dominated, and therefore possibly not habitable.}
\label{tidal_hz}
\end{center} \end{figure}

For illustrative purposes it is helpful to compare the radiative ``habitable" zone (HZ) to the ``tidal" zone, the orbital distance at which tidal dissipation is likely to dominate the internal heat budget of the planet.  {  For this comparison we compute the tidal heat flow using (\ref{q_tidal}) for an Earth-sized planet with $e=0.1$ and $-\mbox{Im}(k_2)=3\times10^{-3}$, similar to present-day Earth.}
Figure \ref{tidal_hz} shows that for $M_{star}<0.3~M_{*}$ the HZ overlaps with the tidal zone, {  defined as when tidal heating is the dominant term in the heat budget}. This diagram implies that Earth-mass planets in the HZ of low mass stars could experience extreme tidal heating and a rapid resurfacing rate that may render the surface uninhabitable.  The HZ and tidal zone intersect because stellar radiation flux is more sensitive to stellar mass than the gravitational tidal potential.  We note that larger eccentricity or tidal dissipation efficiency ($-\mbox{Im}(k_2)$) would push the tidal zone limits out to larger orbital distances, rendering the HZ tidally dominated for larger mass stars.

The one-dimensional dissipation model in (\ref{q_tidal}) assumes a homogeneous body with uniform stiffness and viscosity.  
To derive the dissipation efficiency ($-$Im$(k_2)$) we first define the love number,
\be k_2=\frac{3/2}{1+\frac{19}{2}\frac{\mu}{\beta_{st}} }  \label{k2}  \ee
where $\mu$ is shear modulus and $\beta_{st}$ is effective stiffness \citep{peale1978}.  Writing shear modulus as a complex number and using the constitutive relation for a Maxwell body \citep{henning2009}, one can derive the dissipation efficiency in (\ref{q_tidal}) as
\be -\mbox{Im}(k_2)=\frac{57 \eta \omega}{4 \beta_{st}\left(1+\left[\left(1+\frac{19\mu}{2\beta_{st}}\right) \frac{\eta\omega}{\mu}\right]^2 \right)} \label{im_k2} \ee
where $\eta$ is dynamic viscosity.  We note that this model does not involve a tidal $\tidalq$ factor, rather the rheological response of the mantle is described entirely by $\mbox{Im}(k_2)$.
For comparison with other models, one can compute the standard tidal $\tidalq$ factor of the Maxwell model as
\be  \tidalq =\frac{\eta \omega}{\mu} \label{tidal_q} ~. \ee
The common approximation is then $-\mbox{Im}(k_2)\approx \mbox{Re}(k_2)/\tidalq$ \citep[e.g.][]{goldreich1966,jackson2009,barnes2013}.
Although the tidal $\tidalq$ factor does not appear explicitly in the calculations below, it will be used to calibrate this model to Earth and is useful in comparing it to other models (also see Appendix \ref{appendix_dissipation}).

Following previous studies \citep{sotin2009,behounkova2010,behounkova2011}, we ensure that the model reproduces the observed tidal dissipation in the solid Earth by calibrating the effective material properties in (\ref{im_k2}) appropriately.  This calibration allows us to approximate the total tidal dissipation over the whole mantle by a single volume averaged dissipation function.  
The $\tidalq$ factor of the solid Earth has been estimated empirically to be $\tidalq_E\approx100$ \citep{ray2012}.
Effective viscosity follows an Arrhenius Law form,
\be
\nu=\nu_{ref}\mbox{exp}\left(\frac{E_\nu}{R_g T_m}\right)/\epsilon_{phase}  
\label{nu} \ee
where $\nu=\eta/\rho_m$ is kinematic viscosity, $\rho_m$ is mantle density, $\nu_{ref}$ is a reference viscosity, $E_\nu$ is the viscosity activation energy, $R_g$ is the gas constant, $T_m$ is average mantle temperature, and $\epsilon_{phase}$ accounts for the effect of the solid to liquid phase change (see Table \ref{table1} for a list of constants).  Shear modulus is similarly described,
\be 
\mu=\mu_{ref}\mbox{exp}\left(\frac{E_\mu}{R_g T_m}\right)/\epsilon_{phase}~.
\label{mu}\ee
This model predicts the rapid drop in shear modulus with melt fraction demonstrated experimentally by \citet{jackson2004}.
The reference shear modulus $\mu_{ref}=6.24\times10^4$ Pa and effective stiffness $\beta_{st}=1.71\times10^4$ GPa are calibrated by $k_2=0.3$ and $\tidalq=100$ for the present-day mantle.

We model the influence of melt fraction $\phi$ on viscosity following the parameterization of \citet{costa2009},
\begin{eqnarray}
\epsilon_{phase}(\phi) &=& \frac{1+\Phi^{\delta_{ph}}}{ \left[ 1- F \right]^{B \phi^*} } \label{epsilon_phase}  \\
F &=&  (1-\xi) \mbox{erf} \left[ \frac{\sqrt{\pi}}{2(1-\xi)} \Phi(1+\Phi^{\gamma_{ph}}) \right]    \label{epsilon_phase2}
\end{eqnarray}
where $\Phi=\phi/\phi^*$, and $\phi^*$, $\xi$, $\gamma_{ph}$, and $\delta_{ph}$ are empirical constants (Table \ref{table1}).

\begin{figure} \begin{center}
\includegraphics[width=0.9\linewidth]{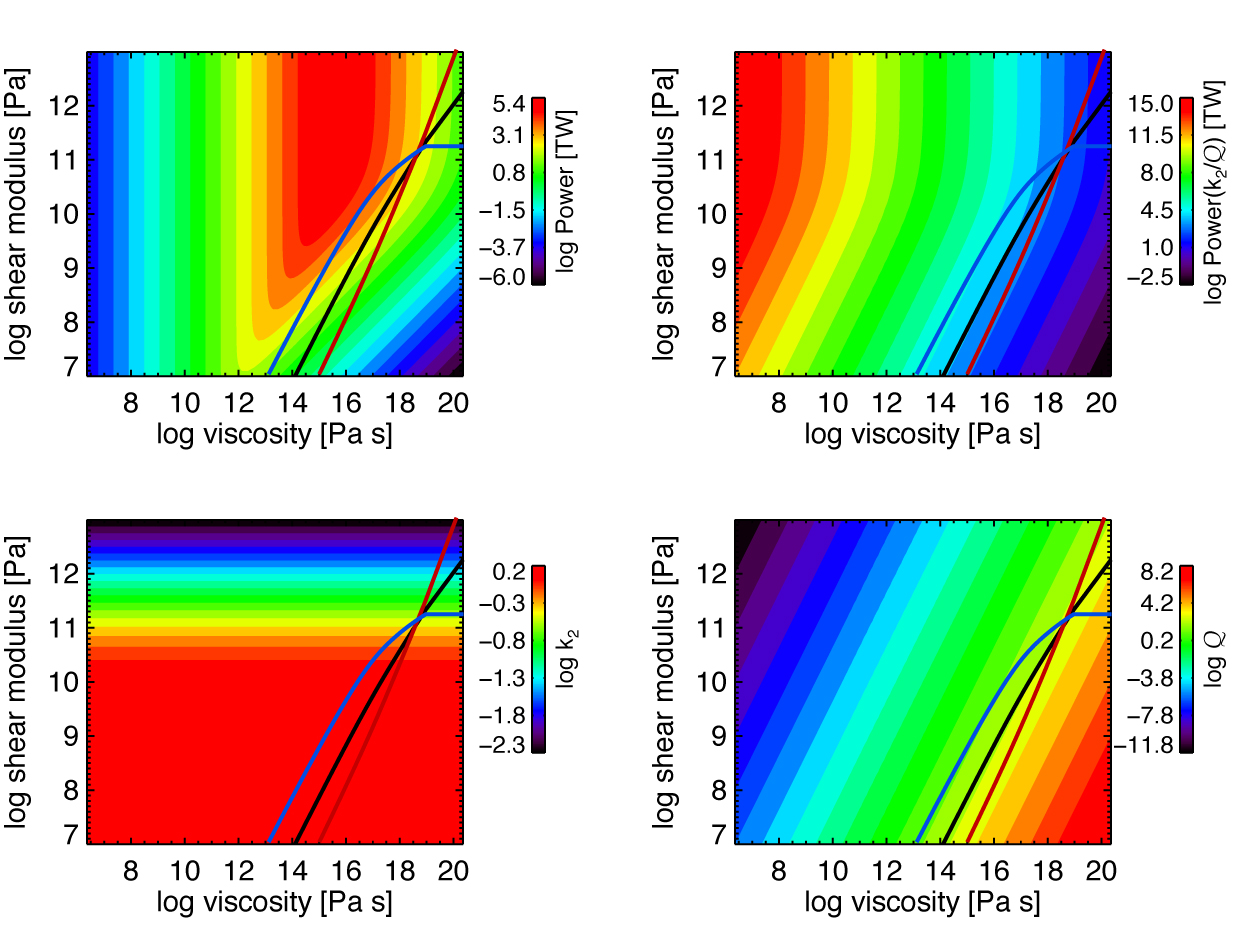}
\put(-425,305){(a)}
\put(-205,305){(b)}
\put(-425,145){(c)}
\put(-205,145){(d)}
\caption{Comparison of tidal properties as a function of viscosity and shear modulus for three shear modulus activation energies $A_\mu=0~\jmol$ (blue line), $2\times10^5~\jmol$ (black line), and $4\times10^5~\jmol$ (red line).  Lines show tracks of $\nu(T_m)$ and $\mu(T_m)$ for mantle temperatures in the range $1500-2000$ K.  (a) Contour of tidal heat flow $Q_{tidal}$.  (b) Contour of tidal power using the approximation $-\mbox{Im}(k_2)=k_2/\mathcal{Q}$.  (c) Contour of Love number $k_2$.  (d) Contour of tidal dissipation factor $\mathcal{Q}$.  Calculations use $M_*=0.1M_{sun}$, $A_\nu=3\times10^5~\jmol$, $e=0.1$.}
\label{contour_mu_nu}
\end{center} \end{figure}

The functional form of tidal shear modulus at high temperature--pressure is not well known, so we investigate the influence of shear modulus activation energy $A_\mu$ on the model.  Contoured in Figure \ref{contour_mu_nu} are tidal power using (\ref{q_tidal}-\ref{im_k2}), tidal power using the approximation $-Im(k_2)= k_2/\mathcal{Q}$, love number $k_2$, and tidal factor $\mathcal{Q}$ as functions of shear modulus $\mu$ and viscosity $\nu$.  Evolution paths of $Q_{tidal}$ as a function of $T_m$ in the range $1500-2000$ K are also shown in these contours for three shear modulus activation energies: $A_{\mu}=0$, $2\times10^5$, and $4\times10^5~\jmol$.  For the nominal activation energy of $A_\mu=2\times10^5~\jmol$ the mantle cooling path passes through a maximum tidal dissipation around $\nu=10^{16}~\mbox{Pa~s}$ and $\mu=10^{10}~\mbox{Pa}$, corresponding to $T_m\approx1800$ K where the mantle is in a visco-elastic state.  In this state the short term tidal response of the mantle is elastic while the long-term response is viscous (see \S \ref{section_ss}).  We note that although these paths pass through a local maximum in $Q_{tidal}$ they do not pass through the global maximum (dark red), which has been invoked for Io \citep{segatz1988}.
The main influence of increasing $A_\mu$ is to shift the dissipation peak to lower temperatures.  The nominal value of $A_\mu=2\times10^5~\jmol$ produces a dissipation peak when melt fraction is about $50\%$.

The tidal dissipation factor $\mathcal{Q}$ (Figure \ref{contour_mu_nu}d) and tidal power using $\mathcal{Q}$ (Figure \ref{contour_mu_nu}b)  change by about one order of magnitude over the entire temperature and activation energy range (Figure \ref{contour_mu_nu}d).  However, tidal power using the Maxwell model in (\ref{q_tidal}) fluctuates by $10^5$ TW over the same range (Figure \ref{contour_mu_nu}a).  
This comparison emphasizes that, although $\mathcal{Q}$ is not far from constant, dissipation using the Maxwell model is significantly different than the $\mathcal{Q}$ approximation (also see Appendix \ref{appendix_dissipation}).  
The Love number $k_2$ increases monotonically with $T_m$ for all cases (Figure \ref{contour_mu_nu}c) up to the limit of $k_2=3/2$ when $\mu/\beta<<2/19$ in (\ref{k2}).

We note that dissipation in this model is a lower bound as dissipation in the liquid is not included, which can occur by resonant dissipation \citep[e.g.][]{tyler2014,matsuyama2014}.  Dissipation in the liquid is not likely to be a major heat source, but could drive mechanical flows in the core \citep{zimmerman2014,lebars2015} and amplify dynamo action there \citep{dwyer2011,mcwilliams2012}.

%%%%%%  THERMAL EVO MODEL %%%%%%%%%%%%%%%%%%%%%%%
\subsection{Thermal Evolution Model}  \label{section_thermal_model}
The thermal evolution of the interior solves the balance of heat sources and sinks in the mantle and core.  The thermal evolution is modeled as in \citet{driscoll2014}, with an updated mantle solidus and inclusion of latent heat release due to magma ocean solidification (see Appendix \ref{appendix_solidus}).
The conservation of energy in the mantle is,
\be Q_{surf}=Q_{conv}+Q_{melt} =Q_{rad}+Q_{cmb}+Q_{man}+Q_{tidal}+Q_{L,man} 
\label{mantle_energy} \ee
where $Q_{surf}$ is the total mantle surface heat flow (in W), $Q_{conv}$ is heat conducted through the lithospheric thermal boundary layer that is supplied by mantle convection, $Q_{melt}$ is heat loss due to the eruption of upwelling mantle melt at the surface, $Q_{rad}$ is heat generated by radioactive decay in the mantle, $Q_{cmb}$ is heat lost from the core across the core-mantle boundary (CMB), $Q_{man}$ is the secular heat lost from the mantle, $Q_{tidal}$ is heat generated in the mantle by tidal dissipation, and $Q_{L,man}$ is latent heat released by the solidification of the mantle.  Crustal heat sources have been excluded because they do not contribute to the mantle heat budget.  Note that heat can be released from the mantle in two ways: via conduction through the upper mantle thermal boundary layer ($Q_{conv}$) and by melt eruption ($Q_{melt}$).  Detailed expressions for heat flows and temperature profiles as functions of mantle and core properties are given in Appendix \ref{appendix_thermalmodel}.

The thermal evolution model assumes a mobile-lid style of mantle heat loss where the mantle thermal boundary layers maintain Rayleigh numbers that are critical for convection.  In contrast, a stagnant lid mantle parameterization would have a lower heat flow than a mobile-lid at the same temperature \citep[e.g.][]{solomatov2000}.  However, a stagnant lid mantle that erupts melt efficiently to the surface can lose heat as efficiently as a mobile-lid mantle with no melt heat loss \citep{moore2013,driscoll2014}.  Io is an example of this style of mantle cooling \citep{oreilly1981,moore2007io}.

Similarly, the thermal evolution of the core is governed by the conservation of energy in the core,
\be Q_{cmb}=Q_{core}+Q_{icb}+Q_{rad,core}   \label{core_energy} \ee
where $Q_{core}$ is core secular cooling, $Q_{rad,core}$ is radiogenic heat production in the core, and heat released by the solidification of the inner core is $Q_{icb}=\dot{M}_{ic}(L_{icb}+E_{icb})$, where $\dot{M}_{ic}$ is the change in inner core mass $M_{ic}$, and $L_{icb}$ and $E_{icb}$ are the latent and gravitational energy released per unit mass at the inner-core boundary (ICB).  

The internal thermal evolution equations are derived by using the secular cooling equation $Q_{i}=-c_i M_i \dot{T}_i$, where $c$ is specific heat and $i$ refers to either mantle or core, in equations (\ref{mantle_energy}) and (\ref{core_energy}).  Solving for $\dot{T}_m$ and $\dot{T}_c$ gives the mantle and core thermal evolution equations,
\begin{equation}
 \dot{T}_m=\left( Q_{cmb} + Q_{rad} +Q_{tidal}+Q_{L,man} -Q_{conv}-Q_{melt} \right)  / M_m c_m  
\label{dot_T_m} \end{equation}
\begin{equation}
\dot{T}_c= -\frac{ (Q_{cmb}-Q_{rad,c})} {M_c c_c - A_{ic} \rho_{ic} \epsilon_c \frac{d R_{ic}}{d T_{cmb}} (L_{Fe}+E_G) }
\label{dot_T_c} \end{equation}
where the denominator of (\ref{dot_T_c}) is the sum of core specific heat and heat released by the inner core growth, $A_{ic}$ is inner core surface area, $\rho_{ic}$ is inner core density, $\epsilon_c$ is a constant that relates average core temperature to CMB temperature, $dR_{ic}/dT_{cmb}$ is the rate of inner core growth as a function of CMB temperature, and $L_{Fe}$ and $E_G$ are the latent and gravitational energy released at the ICB per unit mass (Table \ref{table1}).  See Appendix \ref{appendix_thermalmodel} and \citet{driscoll2014} for more details.  

%%%%%%  ORBITAL EVO MODEL %%%%%%%%%%%%%%%%%%%%%%%
\subsection{Orbital Evolution Model}  \label{section_orbital_model}
The orbital evolution of the planet's semi-major axis $a$ and eccentricity $e$, assuming no dissipation in the primary body (the star), is \citep{goldreich1966,jackson2009,ferrazmello2008}
\be \dot{e}=\frac{21}{2} Im(k_2) \frac{M_*}{M_p} \left(\frac{R_p}{a}\right)^5 n e  \label{ec_dot} \ee
and
\be \dot{a}=2ea\dot{e}   \label{a_dot} .\ee
Mean motion can be replaced by using $n^2=GM_*/a^3$,
\be \dot{e}=\frac{21}{2} Im(k_2) \frac{M_*^{3/2} G^{1/2} R_p^{5}}{M_p} \frac{e}{a^{13/2}} .  \label{ec_dot} \ee

The differential equations for thermal evolution (\ref{dot_T_m}, \ref{dot_T_c}) and orbital evolution (\ref{ec_dot}, \ref{a_dot}) are solved simultaneously to compute coupled thermal-orbital evolutions.

%%%%%%  STEADY STATES   %%%%%%%%%%%%%%%%%%%%%%%
\section{Steady State Solutions}  \label{section_ss}
\begin{figure} \begin{center}
\includegraphics[width=\textwidth]{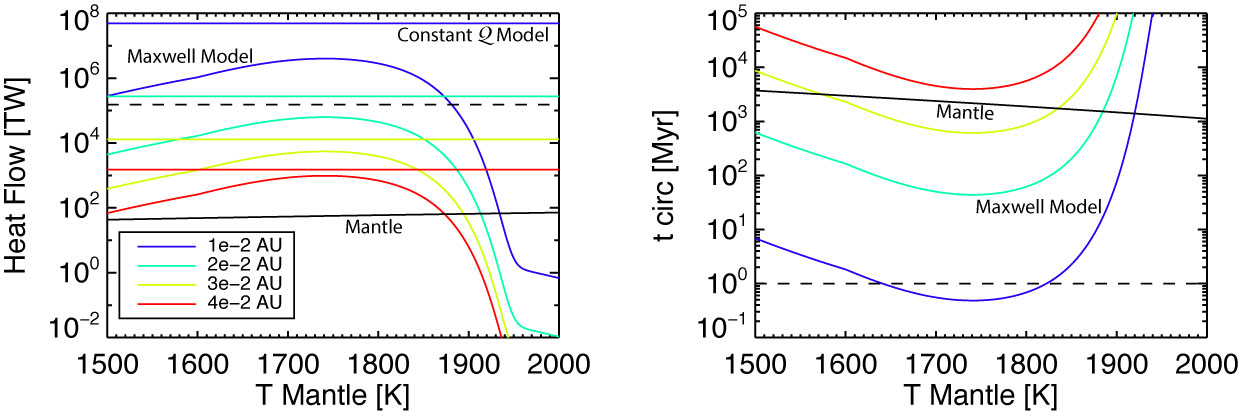}
\put(-475,150){(a)}
\put(-240,150){(b)}
\caption{Tidal dissipation properties as a function of mantle temperature $T_m$ for $M_*=0.1M_{sun}$ and $e=0.1$.  (a)
Comparison of tidal heat flow from the Maxwell model (curves) with the constant $\mathcal{Q}$ model ($\mathcal{Q}=100$, $k_2=0.3$) at four orbital distances (see legend).  Also shown in (a) is the mantle surface heat flow $Q_{surf}$ as a function of temperature (solid black) and constant runaway greenhouse threshold (dashed).
(b) Timescales for orbital circularization using the Maxwell model (same colors as in (a)) and mantle cooling (solid black).   }
\label{plot_ss}%\label{Q_T}
\end{center} \end{figure}

Before exploring the full model it is useful to highlight the influence of tidal heating on the thermal evolution in a steady state sense by comparing heat flows as functions of mantle temperature.  
Figure \ref{plot_ss} shows the tidal heat flow (a) and orbital circularization time (b),
\be t_{circ}=e/\dot{e}  \label{t_circ} \ee 
as a function of mantle temperature for a range of initial orbital distances and eccentricity of $e=0.1$.  
Figure \ref{plot_ss}a shows that a peak in dissipation occurs when the mantle is in a partially liquid visco-elastic state ($T_m\approx1800$ K), where initial tidal perturbations behave elastically and the long timescale relaxation is viscous.  Dissipation is lower in a colder mantle where the response is closer to purely elastic.  Dissipation is also lower in a hotter, mostly liquid mantle where the material behaves viscously, with little resistance to the external forcing.  

Also plotted in Figure \ref{plot_ss}a is the convective mantle cooling curve from (\ref{q_conv}), which reflects the preferred cooling rate of the interior.  Conceptually, a tidal steady state is achieved as the planet cools down from an initially hot state ($T_m>2000$ K) until the convective cooling curve intersects the tidal dissipation (heat source) rate.  This intersection implies that cooling is in balance with tidal heating so that the interior stops cooling.
The steady state occurs around $1850-1950$ K over the range of orbital distances in Figure \ref{plot_ss}a.  The steady state is maintained until the orbit begins to circularize, which drops the dissipation curve and intersection point to lower temperatures.  Circularization continues slowly until the dissipation rate falls below the surface cooling rate, at which point the planet resumes cooling normally with tidal heat playing a minor role in the heat budget.

The time required to circularize, shown in Figure \ref{plot_ss}b, is inversely proportional to dissipation rate through (\ref{ec_dot} and \ref{t_circ}) so that a mantle in a visco-elastic state dissipates orbital energy efficiently and circularizes quickly.  At the inner edge of the habitable zone ($a=0.02$ AU), circularization occurs in less than $\sim1$ Gyr, while on the outer edge circularization requires billions of years or may not occur at all.  {  Also shown in Figure \ref{plot_ss}b is the mantle cooling time $t_{T}=M_m c_m (T_m(0)-T_m)/Q_{surf}$, which is the time required for the mantle to cool from $T_m(0)=2500$ K to $T_m$.  This shows that it takes the mantle $\sim1$ Gyr to adjust to a change in the tidal heat source.
The cooling time is typically longer than the circularization time in the habitable zone (Figure \ref{plot_ss}b), implying that tidal heating can evolve faster than the thermal response of the mantle.}

%%%%%%%%%%%%%%%%%%%%%%%%%%%%%%%%%%%%%%%%%%
%%%%%%  NUMERICAL RESULTS   %%%%%%%%%%%%%%%%%%%%%%%
\section{Model Results: Evolving Orbits}  \label{section_evolvingorbits}
This section presents full thermal-orbital evolutions with a single Earth-mass planet in orbit around a $0.1$ solar mass M star.  
{  We first focus on planets orbiting a $0.1$ solar mass star, where the habitable zone is very close to the star, in order to examine an extreme tidal environment for a lone planet.  Section \ref{section_stellarmass} investigates a range of stellar masses.}
In this section the orbit of the planet is free to evolve according to (\ref{ec_dot} and \ref{a_dot}).  Later in \S \ref{section_fixedorbits} we explore thermal evolutions with fixed orbits.  The models all have $A_\mu=2\times10^5~\jmol$, $T_m(0)=2400$ K,  and $T_c(0)=6000$ K.  The results are independent of initial mantle and core temperatures up to approximately $\pm500$ K.

\subsection{Influence of Initial Orbital Distance}
First, we investigate the evolution of three models that start with $e(0)=0.5$ at three orbital distances: (1) inside the inner edge of the HZ at $a=0.01$ AU, (2) within the HZ at $a=0.02$ AU, and (3) outer edge of HZ at $a=0.05$ AU. 

\begin{figure} \begin{center}
\includegraphics[width=0.8\textwidth]{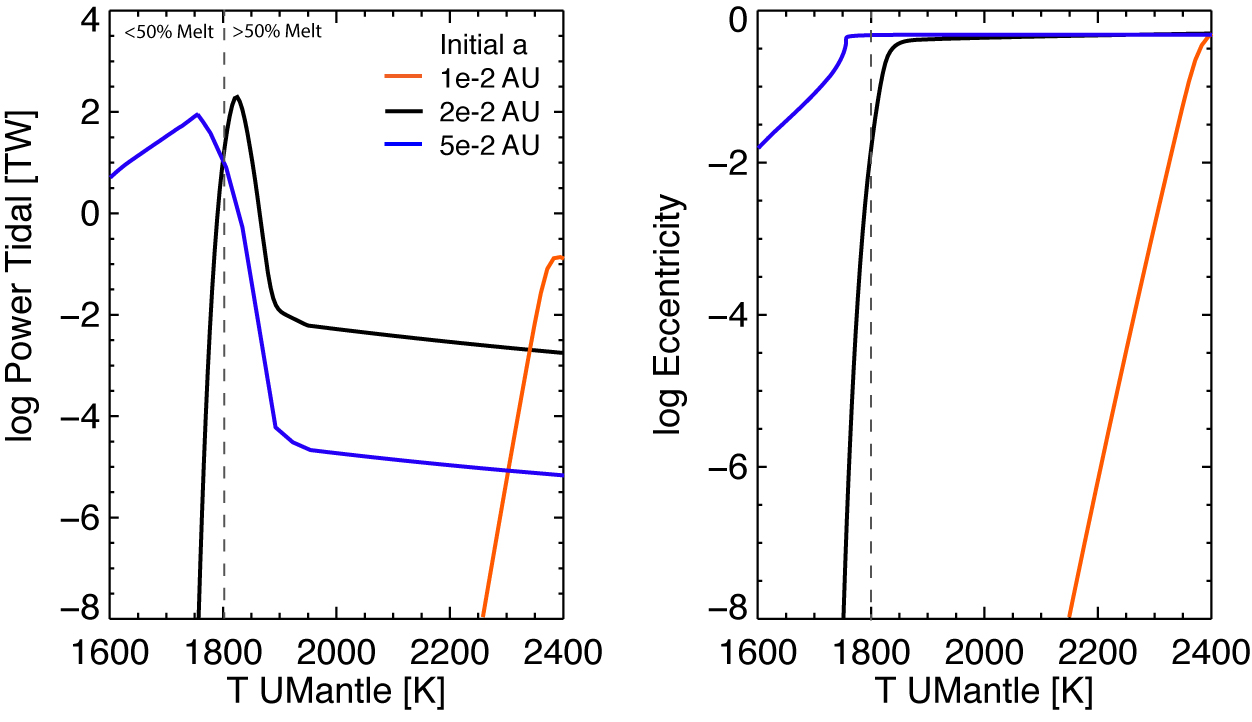}
\caption{Thermal-orbital evolution for three models with initial orbits of $e(0)=0.5$ and $a(0)=0.01$ (red), $0.02$ (black), and $0.05$ (blue).  The temperature that corresponds to 50\% melt fraction is denoted by the dashed line.}
\label{tidal_evo}
\end{center} \end{figure}

Figure \ref{tidal_evo} compares the tidal heat flow and eccentricity of these three models as a function of $T_m$.  The inner case with $a(0)=0.01$ AU begins with a rather high initial tidal heat flow ($Q_{tidal}\sim0.1$ TW) considering the mantle is mostly molten at this time.  Being so close to the star the planet has a fast circularization time (Figure \ref{plot_ss}) so {  eccentricity decreases} rapidly and tidal dissipation effectively ends within 1 Myr (also see Figure \ref{time_evo}c).  {  This implies that circularization occurs during the magma ocean stage.}

The middle case with $a(0)=0.02$ AU begins with a lower tidal heat flow because it orbits farther from the star.  As the mantle cools and solidifies, tidal dissipation evolves through the visco-elastic peak at $T_m\approx1800$ K where the mantle is $\sim50\%$ molten and a peak of $Q_{tidal}\sim100$ TW occurs.  This increase in dissipation drives a rapid circularization (Figure \ref{tidal_evo}b), which then decreases the dissipation as the mantle cools further.

The outer case at $a(0)=0.05$ AU experiences the lowest initial tidal heat flow of $Q_{tidal}(0)\sim10^{-5}$ TW due to it being farthest from the star.  Dissipation remains low and the orbit remains eccentric until the mantle cools to $T_m\sim1800$ K, at which point dissipation increases rapidly.  Tidal heat flow peaks around $Q_{tidal}\sim100$ TW and $T_m\sim1750$ K before decreasing due to decreasing eccentricity.  The peak in dissipation occurs at a slightly lower temperature than the middle case because the eccentricity remains higher longer due to slower circularization.  
In fact, the circularization time is slow enough that after 10 Gyr the model still retains a finite eccentricity of $e\sim0.01$ and a tidal heat flow of $Q_{tidal}\sim10$ TW.  {  This shows that at the outer edge tidal dissipation can linger longer due to slower circularization times.}

A detailed comparison of these three models over time is shown in Figure \ref{time_evo}.  Relatively small differences in their temperature histories (Figure \ref{time_evo}a) are driven by small differences in mantle and core cooling rates (Figure \ref{time_evo}b).  Circularization of the inner model occurs in the first Myr and by 100 Myr for the middle model, while the outer model retains a small eccentricity of $e=2\times10^{-3}$ after 10 Gyr (Figure \ref{time_evo}d).  These circularization times are reflected in the tidal heat flow peaks (Figure \ref{time_evo}b), which occur around 0.1 Myr for the inner case, 10 Myr for the mid case, and 1 Gyr for the outer case.  
{  Inward migration by $10-20\%$ also accompanies this circularization (Figure \ref{time_evo}c).}

%%%  FIG

The thermal evolutions are mainly controlled by secular cooling and radiogenic decay, with tidal dissipation as a temporary energy source.  Mantle heating due to latent heat released during the solidification of the mantle is of order $10^4$ TW until $\sim1$ Myr, then drops below $\sim1$ TW once the mantle is mostly solid around $0.1$ Gyr.  This decrease in latent heat causes the mantle heat flow to drop rapidly between $1-10$ Myr.  Mantle solidification, and the drop in mantle heat flow, occurs slightly later for the inner model for two reasons: (1) the surface heat flow is lower than the other models in the first Myr because the surface is hot, decreasing the upper mantle temperature jump; (2) tidal heating is initially moderate ($\sim0.1$ TW) despite the mantle being mostly molten due to proximity to the star.

These mobile-lid Earth-like models have a strong temperature feedback, or thermostat effect, so that if mantle temperature increases (for example due to tidal dissipation) the viscosity decreases rapidly and the boundary layers thin out, resulting in an increase in the boundary heat flows.  Consequently, increases in internal heat sources are accommodated by increases in heat flows so that the mantle and core cool monotonically.  One minor exception is the brief heating of the core at 1 Gyr due to early radioactive decay in the core.
In contrast, a stagnant lid parameterization with a weaker heat flow-temperature feedback would force the mantle to maintain higher temperatures in order to accommodate the same cooling rates \citep[e.g.][]{solomatov2000,driscoll2014}.  {  Therefore, we expect mobile-lid planets to cool faster, dissipate tidal energy more efficiently, and circularize faster than stagnant-lid planets.  Stagnant lid planets that are strongly tidally heated likely rely on melting rather than conduction to remove heat from the interior, as Io demonstrates today.}

 Core cooling rates are similar at these three orbital distances, which results in similar magnetic moment histories and inner core nucleation times (Figure \ref{time_evo}e).  Inner core nucleation induces a kink in the core compositional buoyancy flux and magnetic moment around 4 Gyr, similar to predictions for the Earth \citep{driscoll2014}.  Surface melt eruption rate is determined by the mantle cooling rate through the upper mantle geothermal gradient, so that the eruption rates at these three orbital distances are similar and follow the mantle heat flow history (Figure \ref{time_evo}f).  After $6$ Gyr the middle and outer planet's mantles are completely solid so that melt eruption ends, while melt eruption continues longer for the inner case due to a slightly hotter mantle.

In summary, Earth-like planets near the inner edge of the HZ around $0.1M_{*}$ stars circularize rapidly (within a few Myr), allowing internal cooling and core dynamo action to proceed similar to Earth.  On the outer edge of the HZ orbital circularization is slower, leading to a prolonged period of tidal dissipation that is accentuated by the cooling of the mantle through a visco-elastic state after $\sim1$ Gyr.  Despite these differences in the tidal evolution, the magnetic and magmatic evolutions of these mobile-lid planets are similar.  These three cases with high initial eccentricities of $e(0)=0.5$ demonstrate the potential for a strong coupling between orbital and thermal evolution by tidal dissipation.

%%%%%  subsection SUMMARY EVO CONTOURS  %%%%%%%
\subsection{Summary Evolution Contours}  \label{section_contours}
\begin{figure} \begin{center}
\includegraphics[width=\textwidth]{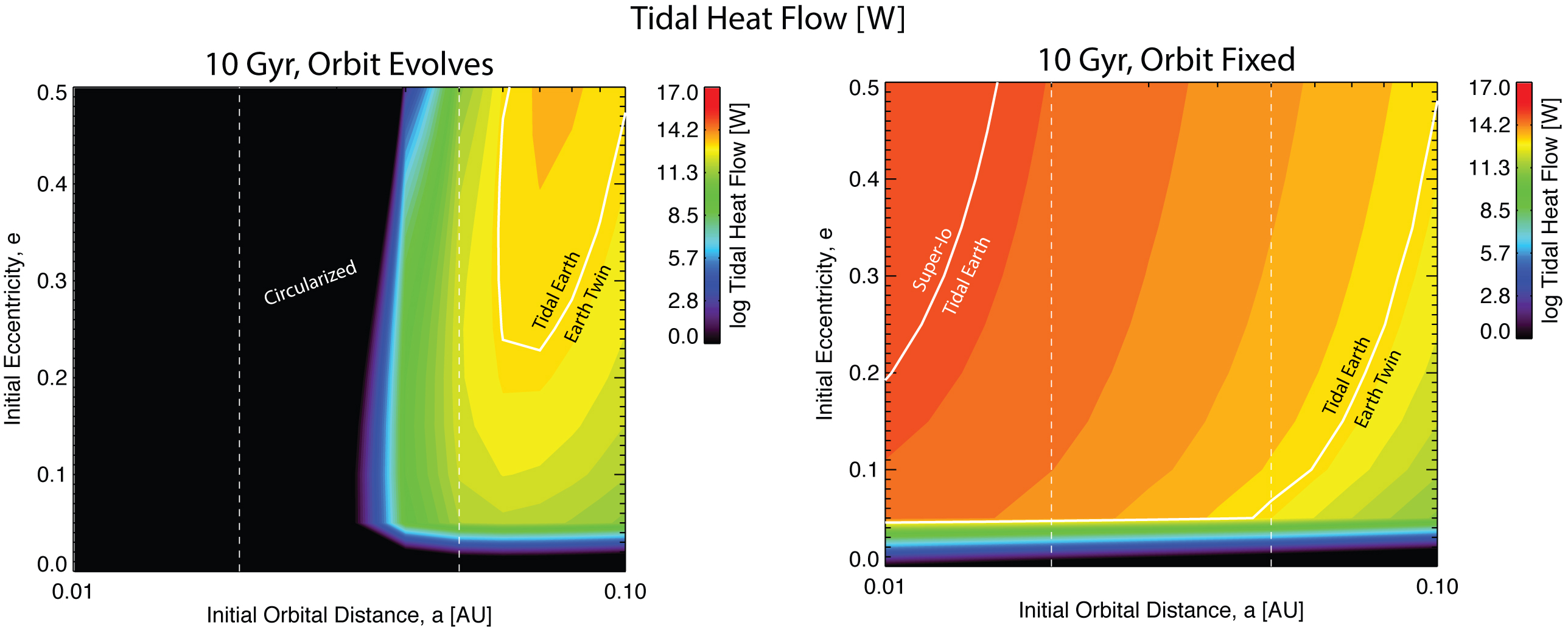}
\put(-480,160){(a)}
\put(-240,160){(b)}
\caption{Contour of (log) tidal heat flow after 10 Gyr for a range of initial orbital distances and eccentricities.  (a) Orbit evolves.  (b) Orbit is fixed.  The tidal heat flow boundaries defined by \citet{barnes2013b} are shown for Earth Twins $Q_{tidal}<20$ TW, Tidal Earths $20<Q_{tidal}<1020$ TW, and Super-Io's for $Q_{tidal}>1020$ TW.}
\label{contour_hftidal}
\end{center} \end{figure}

\begin{figure} \begin{center}
\includegraphics[width=\textwidth]{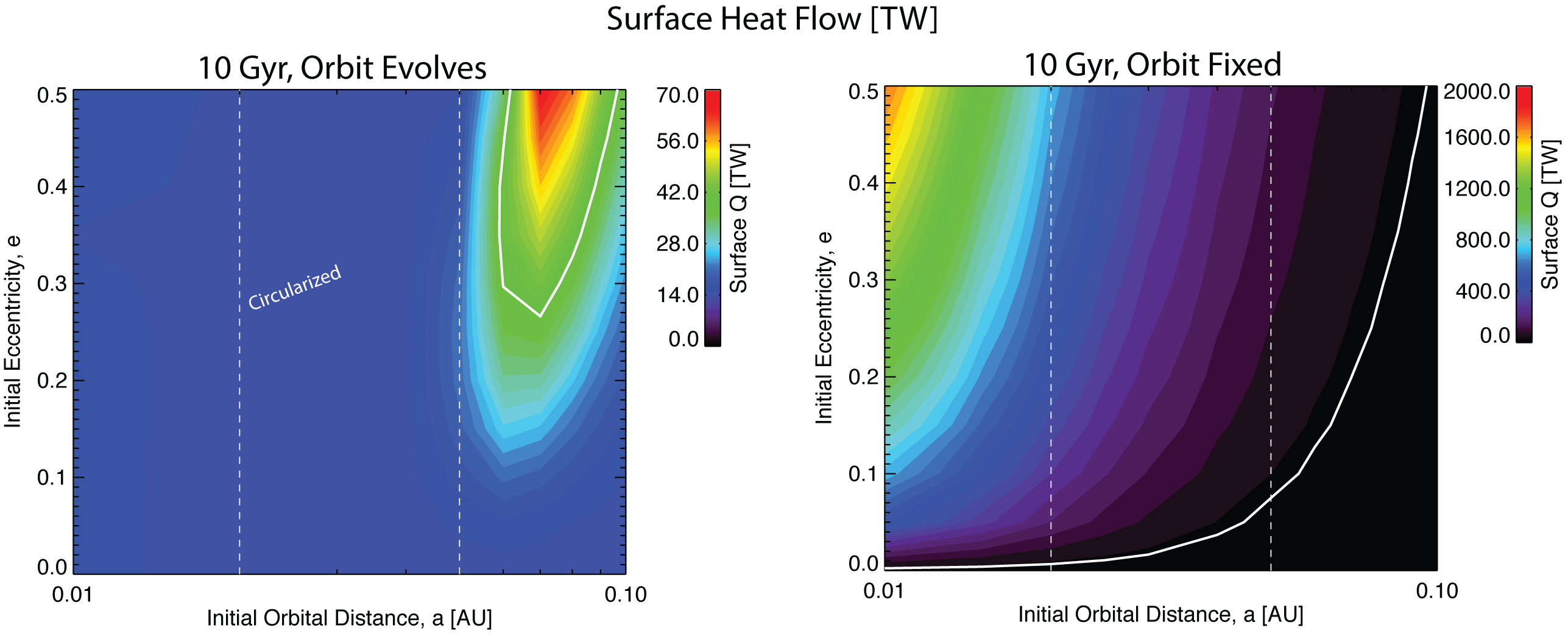}
\put(-480,160){(a)}
\put(-240,160){(b)}
\caption{Contour of surface heat flow after 10 Gyr for a range of initial orbital distances and eccentricities.  (a) Orbit evolves.  (b) Orbit is fixed.  White contour line shown at Earth's present-day surface heat flow ($Q_{surf}^*=40$ TW).
}
\label{contour_hfsurf}
\end{center} \end{figure}

\begin{figure} \begin{center}
\includegraphics[width=\textwidth]{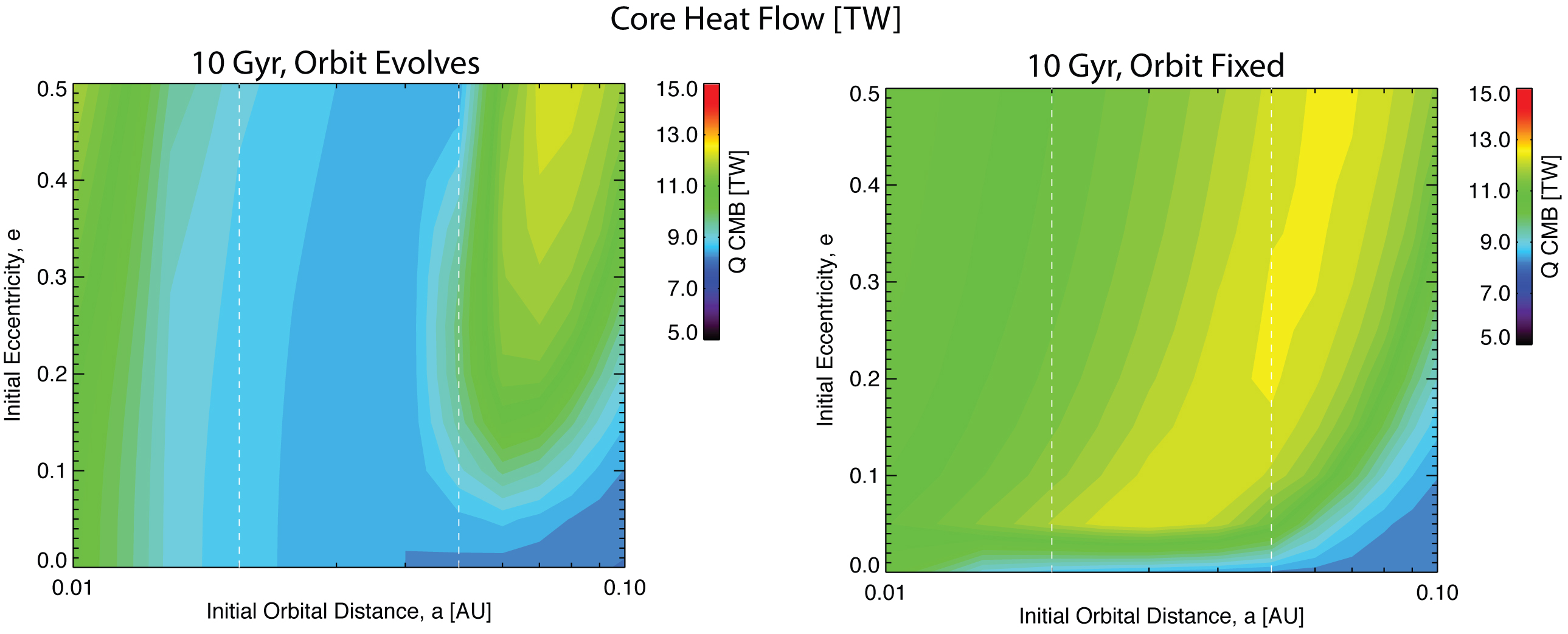}
\put(-480,160){(a)}
\put(-240,160){(b)}
\caption{Contour of core heat flow after 10 Gyr for a range of initial orbital distances and eccentricities.  (a) Orbit evolves.  (b) Orbit is fixed.}
\label{contour_hfcmb}
\end{center} \end{figure}

\begin{figure} \begin{center}
\includegraphics[width=\textwidth]{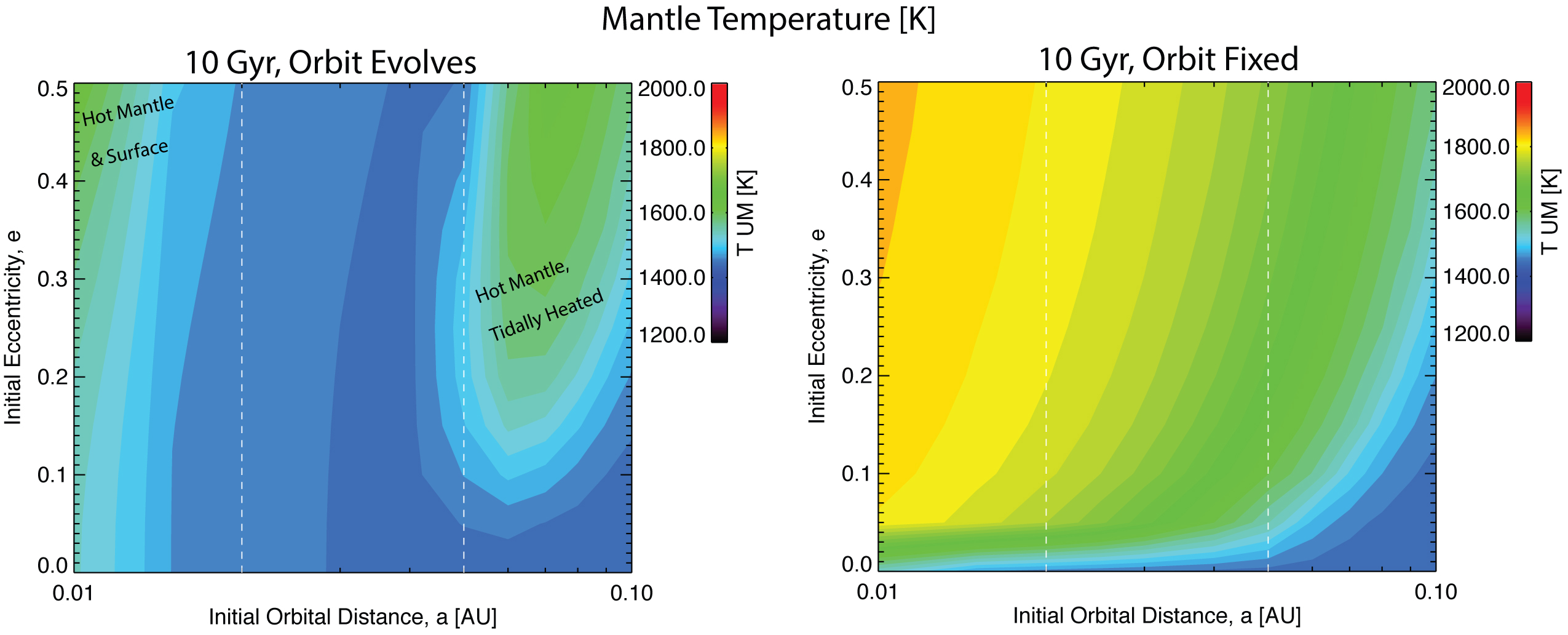}
\put(-480,160){(a)}
\put(-240,160){(b)}
\caption{Contour of mantle temperature after 10 Gyr for a range of initial orbital distances and eccentricities.  (a) Orbit evolves.  (b) Orbit is fixed.}
\label{contour_tman}
\end{center} \end{figure}

\begin{figure} \begin{center}
\includegraphics[width=\textwidth]{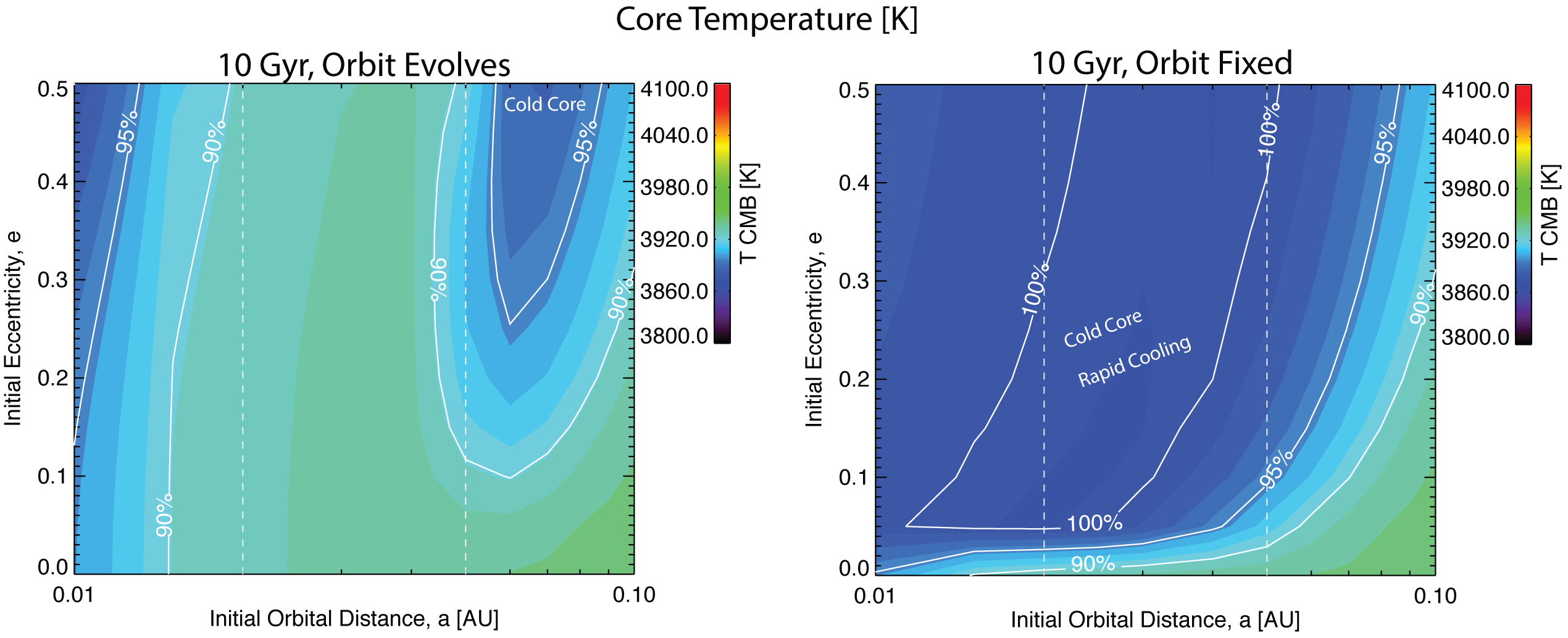}
\put(-480,160){(a)}
\put(-240,160){(b)}
\caption{Contour of core temperature after 10 Gyr for a range of initial orbital distances and eccentricities.  Line contours show solid core fraction $R_{ic}/R_c$ as a percentage (i.e.\ 100\% corresponds to a completely solid core).
(a) Orbit evolves.  (b) Orbit is fixed.}
\label{contour_tcmb}
\end{center} \end{figure}

\begin{figure} \begin{center}
\includegraphics[width=\textwidth]{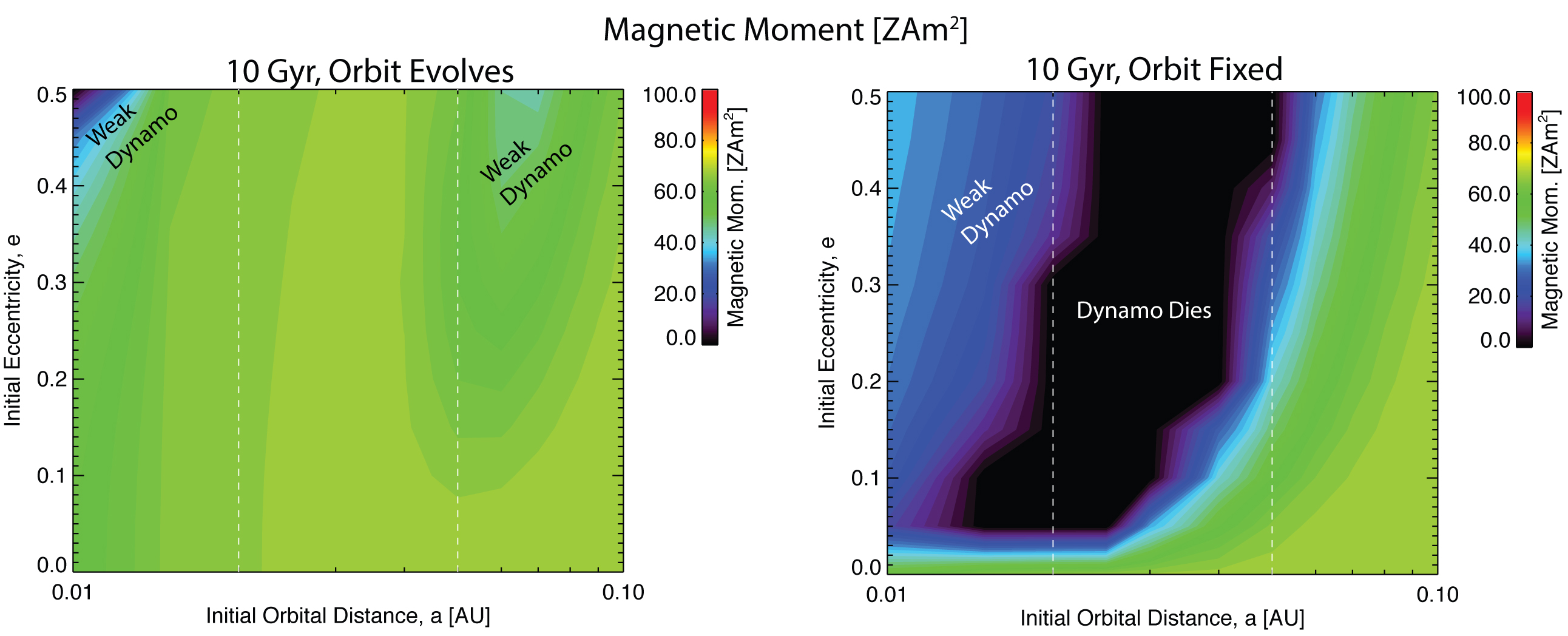}
\put(-480,160){(a)}
\put(-240,160){(b)}
\caption{Contour of magnetic moment after 10 Gyr for a range of initial orbital distances and eccentricities.  (a) Orbit evolves.  (b) Orbit is fixed.   For reference, Earth's present day magnetic moment is about $80$ ZAm$^2$.}
\label{contour_mm}
\end{center} \end{figure}

\begin{figure} \begin{center}
\includegraphics[width=\textwidth]{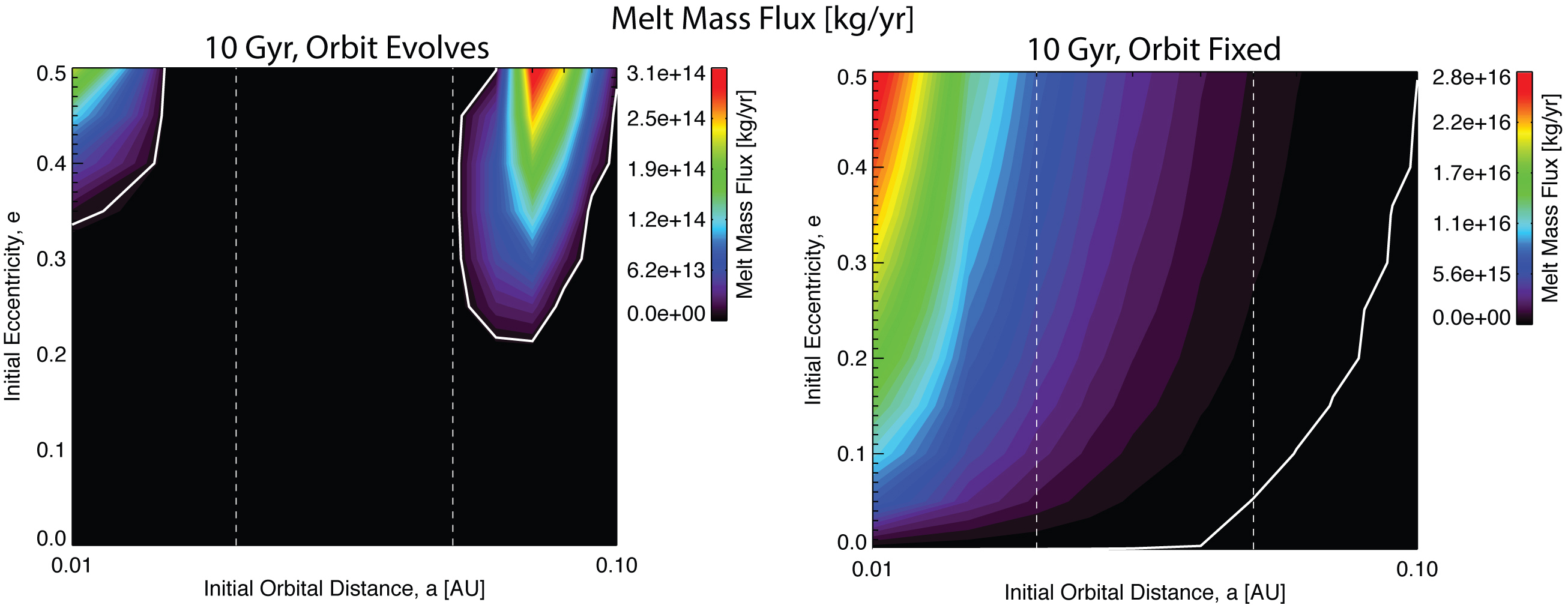}
\put(-480,160){(a)}
\put(-240,160){(b)}
\caption{Contour of surface melt mass flux after 10 Gyr for a range of initial orbital distances and eccentricities.  (a) Orbit evolves.  (b) Orbit is fixed.  White line contour denotes Earth's approximate present-day mid-ocean ridge melt flux ($10^{13}~\mbox{kg~yr}^{-1}$).  Note color scales in (a) and (b) are different.}
\label{contour_massflux}
\end{center} \end{figure}

In this section we compare the final states (after 10 Gyr) of orbital-thermal evolution for 132 models that span a range of initial orbital distances of $0.01-0.10$ AU and initial eccentricities of $0-0.5$.  The results are displayed as contours in initial orbital $a-e$ space (Figures \ref{contour_hftidal}-\ref{contour_massflux}).
Here we consider models whose orbits evolve (left panels of Figures \ref{contour_hftidal}-\ref{contour_massflux}), while \S \ref{section_fixedorbits} below considers models whose orbits are fixed (right panels of Figures \ref{contour_hftidal}-\ref{contour_massflux}).

\begin{figure} \begin{center}
\includegraphics[width=\textwidth]{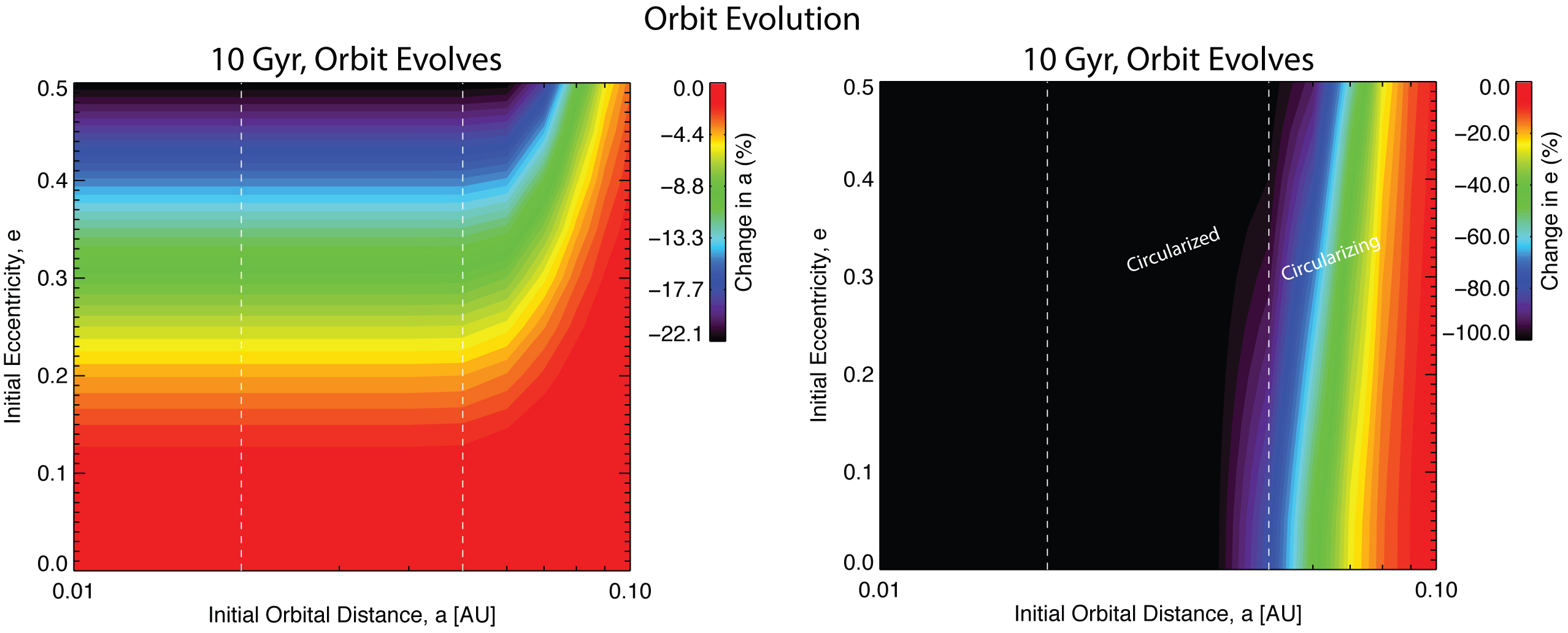}
\put(-480,160){(a)}
\put(-240,160){(b)}
\caption{Contour of orbital evolution after 10 Gyr for a range of initial orbital distances and eccentricities.  (a) Change in orbital distance: $(a-a_0)/a_0$.  (b) Change in eccentricity: $(e-e_0)/e_0$.  Orbits are free to evolve in both panels.  The habitable zone in denoted by vertical dashed white lines.}
\label{contour_orbitevo}
\end{center} \end{figure}

Figure \ref{contour_orbitevo} shows the fractional change in orbital distance (a) and eccentricity (b).  After 10 Gyr most models have circularized within the HZ due to tidal dissipation (Figure \ref{contour_orbitevo}b).  The iso-contour lines in Figure \ref{contour_orbitevo}b are nearly vertical because eccentricity evolution is proportional to $e/a^{13/2}$ in (\ref{ec_dot}).  In other words, orbital circularization is a stronger function of orbital distance than eccentricity.  Circularization also causes the orbit to migrate inwards (Figure \ref{contour_orbitevo}a), although this results in a maximum inwards migration of only $22\%$ of the initial distance.  
The evolution of orbital distance, which produces mainly horizontal iso-contours (Figure \ref{contour_orbitevo}a), is controlled by initial eccentricity because migration is proportional to $e^2/a^{11/2}$ by (\ref{a_dot}); hence migration ($\dot{a}$) is a stronger function of eccentricity than circularization ($\dot{e}$).

%% FIG

Figure \ref{contour_hftidal}a contours tidal heat flow for these models.  {  We identify the tidal heat flow boundaries defined by \citet{barnes2013b} as an Earth Twin for $Q_{tidal}<20$ TW, a Tidal Earth for $20<Q_{tidal}<1020$ TW, and a Super-Io for $Q_{tidal}>1020$ TW.}
Models that circularize by 10 Gyr have zero tidal heating.  At the outer edge of the HZ circularization is still occurring at a rate that is proportional to the change in $e$ in Figure \ref{contour_orbitevo}b.  Beyond $a\sim0.07$ AU tidal dissipation is too weak to result in any signifiant circularization.  Therefore, there are gradients in $Q_{tidal}$ on both sides of this boundary at $a\sim0.07$.  There is also a decrease in $Q_{tidal}$ with initial $e$ because models with low initial $e$ circularize earlier.  The combination of these effects results in a peak in tidal dissipation around $a\sim0.07$ AU and $e\sim0.5$.

%%%FIG

This peak in tidal heat flow causes a slight increase in heat flow (Figure \ref{contour_hfsurf}a) and mantle temperature (Figure \ref{contour_tman}a).  This is an example of the steady state behavior of Figure \ref{plot_ss} where higher tidal heat flows intersect the convective heat flow curve at higher mantle temperatures.  Hotter mantle temperatures also cause the lower mantle to have lower viscosity,  thinner boundary layers, and increased core heat flows (Figure \ref{contour_hfcmb}a).
A second maximum in core heat flow occurs at the innermost orbits due to the high surface temperature insulating the mantle, which keeps mantle temperature high (Figure \ref{contour_tman}a) and thins the lower mantle thermal boundary layer.  Core temperature is low where core heat flow is high (Figure \ref{contour_tcmb}a) due to secular cooling of the core.

After 10 Gyr of significant core cooling all models have a large solid inner core.  The size of the inner core (also contoured in (Figure \ref{contour_tcmb}) is proportional to $Q_{cmb}$ (Figure \ref{contour_hfcmb}a) and is between $80-100$\% of the core radius (also see Figure \ref{time_evo}e).  Where the core is entirely solid no dynamo action is possible (upper left corner of Figure \ref{contour_mm}a), and where the core is mostly solid the magnetic moment is weak due to the small size of the dynamo region (upper right corner of Figure \ref{contour_mm}a).

The eruption of melt to the surface (Figure \ref{contour_massflux}a) is controlled by the upper mantle geothermal gradient and thus proportional to mantle heat flow with a peak around $0.07$ AU.  A secondary peak in melt mass flux at close-in orbits (upper left corner of Figure \ref{contour_massflux}a) is caused by a slightly higher mantle temperature associated with a hotter, insulating surface (Figure \ref{contour_tman}a).  

%%% FIG

%%%%%%  FIXED ORBIT   %%%%%%%%%%%%%%%%%%%%%%%
\section{Model Results: Fixed Orbits}  \label{section_fixedorbits}
In this section we consider planets whose orbits are fixed ($\dot{e}=\dot{a}=0$).  This includes eccentric orbits, which could be fixed, for example, through interactions with a planetary companion \citep{vanlaerhoven2014}.  Figures \ref{contour_hftidal}b-\ref{contour_massflux}b show contours in orbital space, similar to those discussed above except with fixed orbits.

\subsection{Example of Tidal Steady State}
Figure \ref{fixed_example} shows the time evolution of a specific case with a fixed orbit of $a=0.02$ AU and $e=0.5$ that reaches a tidal steady state, an example of the scenario described in \S \ref{section_ss}.  Tidal heating initially starts low ($\sim10^{-3}$ TW) before increasing as the mantle cools for the first 10 Myr, until the mantle reaches a steady state temperature of $T_m\approx 1800$ K.  In this steady state heat loss is balanced by internal sources so that mantle cooling becomes insignificant.  The steady state surface heat flow ($Q_{surf}\approx1000$ TW) corresponds to a surface heat flux similar to that of Io ($q_{surf}\approx2$ W m$^{-2}$), implying that this planet might be better characterized as a super-Io than Earth-like \citep[e.g.][]{barnes2010}.
The tidal steady state still allows the core to cool slowly because a significant temperature difference between the mantle and core persists.  Core cooling drives a core dynamo for all 10 Gyr, although the magnetic moment rapidly declines as the core is nearly entirely solid by 10 Gyr (Figure \ref{fixed_example}c).

%%% FIG

\subsection{Summary Contours}

These fixed orbit models, in contrast with the evolving models in \S \ref{section_evolvingorbits}, have tidal heat flows that are mainly determined by the orbital state rather than the cooling history.  Specifically, $Q_{tidal}$ increases with $e$ and decreases with $a$, producing a maximum in the upper left corner of Figure \ref{contour_hftidal}b.  Mantle heat flow (Figure \ref{contour_hfsurf}b) and temperature (Figure \ref{contour_tman}b) increase with tidal heating due to the positive feedback between mantle temperature and surface heat flow.

Core heat flow peaks in models at moderate orbital distances where tidal heat flow is similar in magnitude to the sum of all other mantle heat sources; i.e.\ $Q_{tidal}\sim Q_{cmb}+Q_{man}+Q_{rad}$ (Figure \ref{contour_hfcmb}b).  This peak can be understood by considering how $Q_{cmb}$ behaves at the two tidal extremes: (1) where tidal heating is strong (upper left corner of Figure \ref{contour_hftidal}b) the mantle is forced into a hot steady state so that surface heat flow can accommodate all heat sources, which thins the lower mantle thermal boundary layer and allows a moderate core heat flow of $Q_{cmb}\sim10$ TW; (2) where tidal heating is weak (lower right corner of Figure \ref{contour_hftidal}b) the mantle and core are free to cool similar to Earth, so that $Q_{cmb}$ decreases monotonically over time.  
{  In between these limits tidal dissipation heats the mantle slightly, increasing the surface heat flow, but does not dominate the heat budget, which allows the interior to cool.  }
Note that even when mantle temperature is high ($\sim2000$ K) it is still significantly colder than the core ($\sim3800$ K) so that the mantle and core are not in thermal equilibrium and the core is forced to cool. 
The effect is to produce a region where a modest amount of tidal dissipation actually promotes core cooling, similar to the peak in the evolving models (Figure \ref{contour_hfcmb}a).  We refer to this 30\% increase in $Q_{cmb}$ as the super-cooling of the core.  

The influence of this peak in core cooling rate on the dynamics of the core is dramatic.  Intuitively, core temperature is lowest where core cooling is high (center of Figure \ref{contour_tcmb}b), and the coldest models with $T_{cmb}\approx3850$ K have completely solid cores (i.e.\ $R_{ic}=R_{core}$).  A fully solid core prevents fluid motion and therefore dynamo action.  These models lose their dynamos at $\sim9.8$ Gyr, so they have dynamos a vast majority of the time. {  Note that our magnetic scaling law likely provides an upper limit on the dynamo lifetime because the scaling law was derived from thick shell dynamos and does not account for stratified layers.}  

This prediction implies that there is a dip (and possibly a gap) in magnetic field strength for tidally heated and orbitally fixed planets over most of the HZ (Figure \ref{contour_mm}b).  
We note that our core liquidus does not include light element depression \citep[e.g.][]{hirose2013}, which would tend to slow inner core nucleation and allow some of these models to maintain a liquid region slightly longer.  This result emphasizes the difference between core cooling and dynamo action: cooling is ongoing (at least until thermal equilibration), whereas dynamo lifetime, which relies on convection in the liquid, is limited by the solidification time of the core \citep[see also][]{gaidos2010,tachinami2011}.  In other words, rapid core cooling is helpful for temporarily driving dynamo action but shortens the lifetime of the dynamo.

The eruption of mantle melt to the surface follows surface heat flow (Figure \ref{contour_massflux}b).  The extreme mass fluxes of $10^{16}~\mbox{kg~yr}^{-1}$ correspond to a global basalt layer resurfacing rate of $7~\mbox{m~kyr}^{-1}$.  For reference, the Siberian traps, one of the largest igneous provinces on Earth and thought to be responsible for the Permian mass extinction event, is estimated to have produced a basalt layer at a rate of $1~\mbox{m~kyr}^{-1}$ over the area of the traps \citep{reichow2002}.  Therefore, continuous eruption rates of $\sim10^{16}~\mbox{kg~yr}^{-1}$ are likely to prevent such planets from being habitable.

We also compute the same range of models but only fixing eccentricity, allowing $a$ to evolve.  This might occur if a neighboring planet forces the eccentricity but allows inward migration.
In these cases we find that all models with initial orbits of $a<0.05$ AU (or $a<0.02$ AU) and $e>0.2$ (or $e>0.1$) migrate into the central star within 10 Gyr, and most by 5 Gyr.

%%%%%  SECTION: Runaway Greenhouse  %%%%%%%
\section{Internally Driven Runaway Greenhouse} \label{section_runaway}
\begin{figure} \begin{center}
\includegraphics[width=\textwidth]{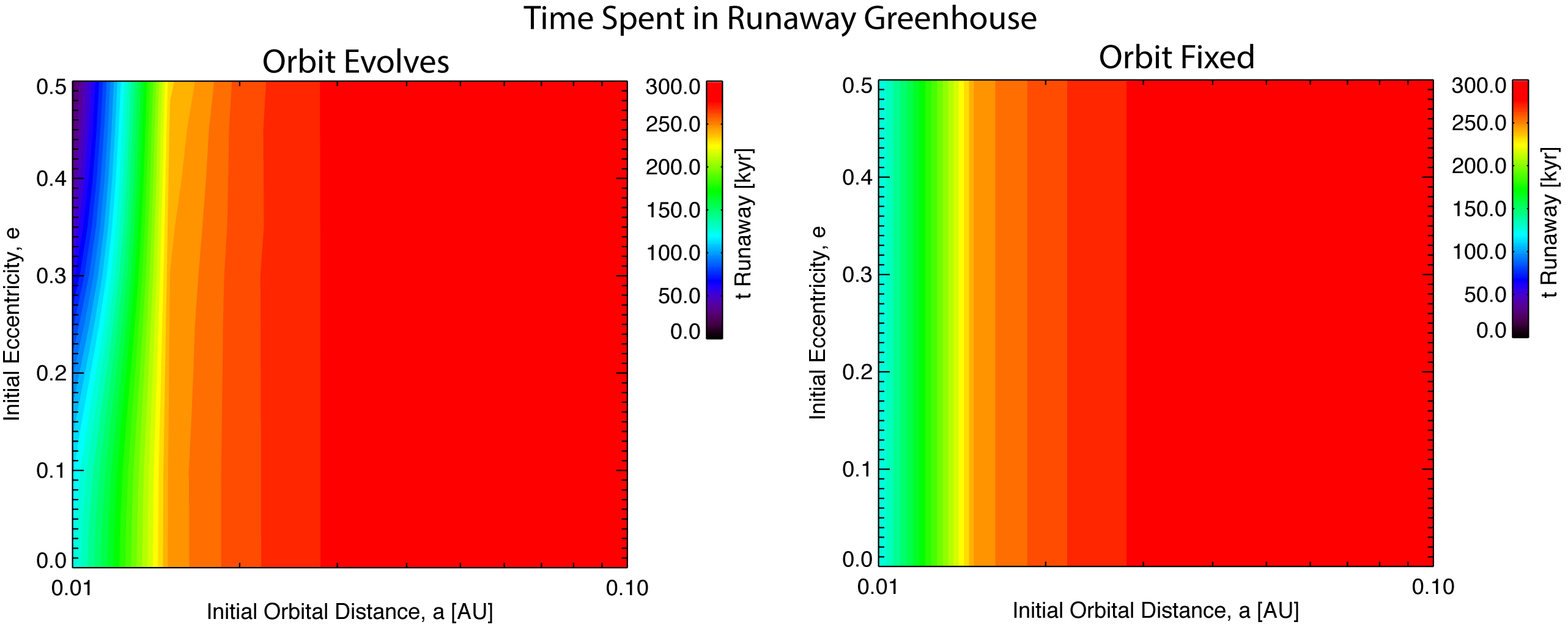}
\put(-480,160){(a)}
\put(-240,160){(b)}
\caption{Contour of time spent in an internally driven runaway greenhouse, defined as when surface heat flow exceeds the threshold for a runaway greenhouse ($300~\mbox{W~m}^{-2}$, or $1.53\times10^{17}$ W), for a range of initial orbital distances and eccentricities.  (a) Orbit evolves.  (b) Orbit is fixed.}
\label{contour_runaway}
\end{center} \end{figure}

{  As described by \citet{barnes2013}, if interior heat flux exceeds the limit at which energy can be radiated from the top of the atmosphere then runaway heating of the surface occurs, evaporating the ocean, and leading to rapid water loss \citep{goldblatt2012}. 
Figure \ref{contour_runaway} shows the time spent in an internally driven runaway greenhouse, defined as the period of time when the surface heat flux exceeds the threshold $q_{runaway}=300~\mbox{W~m}^{-2}$.
 
For both evolving (Figure \ref{contour_runaway}a) and fixed (Figure \ref{contour_runaway}a) orbital models the   runaway greenhouse period is shorter at closer orbital distances, almost independent of eccentricity.  This implies that tides, which depend strongly on eccentricity, play a minor role in the length of the runaway greenhouse state.  The runaway greenhouse state is shorter for close-in planets because they have higher effective surface temperatures closer to the star, which insulates the mantle and decreases the initial surface heat flow.  With lower initial surface heat flows, these inner planets drop below the runaway heat flow threshold earlier (Figure \ref{time_evo}b).  
A second trend in Figure \ref{contour_runaway}a towards even shorter times spent in a runaway greenhouse is found for the inner-most, high eccentricity planets.  This drop in surface heat flow at around $50-100$ kyr occurs during the circularization of the inner planets' orbits, when the tidal heat flow rapidly declines (Figure \ref{time_evo}b).  Circularization causes a small dip in the surface heat flow as the interior temperatures and heat flows adjust to the smaller internal (tidal) heat source.  This adjustment to lower heat flows, although seemingly minor, actually shortens the time spent above the runaway threshold (Figure \ref{time_evo}b).

Interestingly, when the heat flow is high enough to drive a runaway greenhouse during the first few hundred Myr the mantle is so hot that tidal dissipation is inefficient.  Typically tidal dissipation is not a major heat source until the mantle solidifies and cools down to $\sim$1800 K, which occurs after the runaway greenhouse and magma ocean phases.

In summary, we find that mobile lid Earth-like planets typically spend several hundred thousand years in an internally driven atmospheric runaway greenhouse state and that tidal dissipation in the mantle at this time plays a minor role.  
The runaway greenhouse timescale ($\sim100$ kyr) is shorter than the typical magma ocean solidification time ($\sim10$ Myr), a period when the surface is likely uninhabitable anyway.  These models assume mobile lid cooling at all times, however \citet{foley2012} proposed that a runaway greenhouse could induce a transition from mobile to stagnant lid, which would also slow internal cooling and would be detrimental to habitability.
In \S \ref{section_stellarmass} we explore more generally how tides may affect habitability by computing the length of time spent in a tidally dominated state for a range of stellar masses.
}

%%%%%%%%%%   Mstar INNER EDGE CONTOURS  %%%%%%%%%
\section{Influence of Stellar Mass}  \label{section_stellarmass}
\begin{figure} \begin{center}
\includegraphics[width=\textwidth]{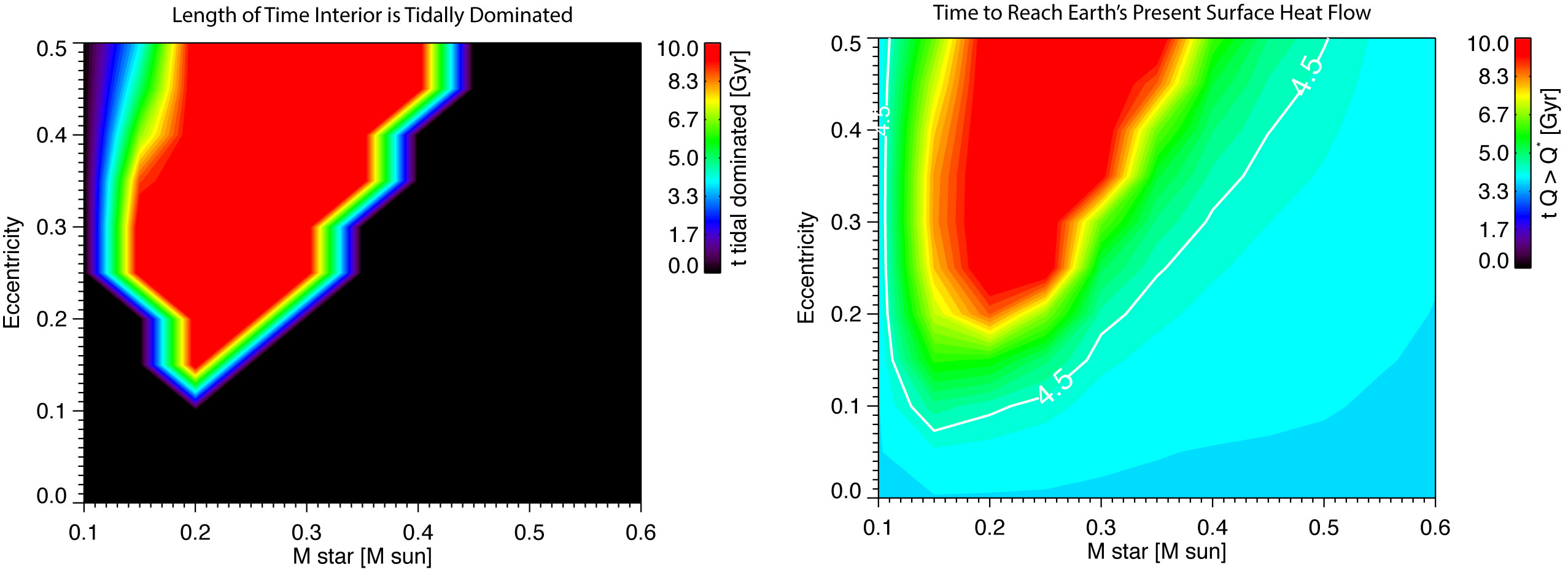}
\put(-480,160){(a)}
\put(-240,160){(b)}
\caption{Contour of (a) time spent in a tidally dominated state (i.e. $Q_{tidal}/Q_{total} \ge 0.5$) and (b) time to reach Earth's present-day surface heat flow ($Q_{surf}=40$ TW).  In (b) a white contour line is shown at $4.5$ Gyr. }
\label{contour_times}
\end{center} \end{figure}

{ 
The above calculations assumed a stellar mass of $0.1~M_{sun}$.  In this section we explore the influence of stellar mass, in the range $0.1-0.6~M_{sun}$, at the inner edge of the habitable zone.  Similar to the contours in \S \ref{section_contours}, we compute a grid of models with a range of initial eccentricities of $0-0.5$ and allow the orbit to evolve in time.  The initial orbital distance is set just outside the inner edge of the radiative habitable zone, which is derived from the stellar mass by the parametric equations of \citet{kopparapu2014}, so that the planet remains in the habitable zone after 10 Gyr of orbital migration.

Figure \ref{contour_times} summarizes the results of these models in terms of two timescales: (a) the time spent in a tidally dominated state, defined as when the tidal heat flow $Q_{tidal}$ is 50\% or more of the total surface heat flow $Q_{surf}$; (b) the time to reach Earth's present-day surface heat flow of $Q_{surf}^*=40$ TW.  

The island-like shapes of these time contours can be explained by a combination of three physical effects.  First, planets with initially low eccentricity ($e(0)<0.1$) experience weak tides and spend little time, if any, in the tidally dominated regime.  At higher eccentricity tides become stronger, so that more eccentric planets are tidally dominated longer (Figure \ref{contour_times}a).
Second, as stellar mass increases the habitable zone moves to larger orbital distances and the tidal dissipation decreases because tidal dissipation in (\ref{q_tidal}) is a stronger function of orbital distance ($\propto a^{-15/2}$) than stellar mass ($\propto M_*^{+5/2}$).  The net result is a decrease in tidal dissipation within the habitable zone for increasing stellar mass, and shorter time spent in the tidally dominated state (Figure \ref{contour_times}a).  This effect produces contour boundaries with positive slope in Figure \ref{contour_times}.
Third, models with high initial eccentricity ($e(0)>0.2$) and close-in initial orbits around low mass stars ($M_{star}<0.12$) experience extreme early tides that drive rapid orbital circularization.  This leads to short times spent in the tidally dominated state.

Figure \ref{contour_times}b, similar to Figure \ref{contour_times}a, shows that eccentric planets on the inner edge around $0.15-0.4~M_{sun}$ stars maintain surface heat flows in excess of $Q_{surf}^*$ for 10 Gyr due to strong tidal dissipation.  Interestingly, Figure \ref{contour_times}b shows that planets that experience only a temporary period of tidal heating actually cool to an Earth-like heat flow before 4.5 Gyr.  These planets cool faster than Earth because their thermal adjustment timescale is longer than their circularization (or tidal heating) timescale, so they are still adjusting to the new heat balance with a lower tidal heat source.  In other words, the surface heat flow, that was increased during the tidal heating phase, is still slightly larger than it would have been with no tidal heating.  This super-cooling effect was also discussed in \S \ref{section_contours}.

In summary, tides are more influential around low mass stars.  For example, planets around $0.2~M_{sun}$ stars with eccentricity of $0.4$ experience a tidal runaway greenhouse for 1 Gyr and would be tidally dominated for 10 Gyr.  These time scales would increase if the orbits were fixed, for example by perturbations by a secondary planetary companion.  We find a threshold at a stellar mass of $0.45M_{sun}$, above which the habitable zone is not tidally dominated.  These stars would be favorable targets in the search for geologically habitable Earth-like planets as they are not overwhelmed by strong tides.
}

%%%%%%  DISCUSSION   %%%%%%%%%%%%%%%%%%%%%%%
\section{Discussion}  \label{section_discussion}
In summary, we have investigated the influence of tidal dissipation on the thermal-orbital evolution of Earth-like planets around {  M-stars with masses $0.1-0.6~M_{sun}$}.  A thermal-orbital steady state is illustrated where, under certain conditions, heat from tidal dissipation is balanced by surface heat flow.  We find that mantle temperatures in this balance are hotter for planets with shorter orbital distances and larger eccentricities.  Orbital energy dissipated as tidal heat in the interior drives both inward migration and circularization, with a circularization time that is inversely proportional to the dissipation rate.  The cooling of an eccentric planet in the habitable zone leads to a peak in the dissipation rate as the mantle passes through a visco-elastic rheology state.  {  Planets around $0.1$ solar mass stars} with initial orbits of $a<0.07$ AU circularize before 10 Gyr, independent of initial eccentricity.  Once circular, these planet cool monotonically and maintain dynamos similar to Earth.  {  Generally, we find that tidal dissipation plays a minor role on the dynamo history if the orbit is free to evolve in time.}

{  When} the orbit is fixed the planet cools until a tidal steady state balance {  between tidal dissipation and surface cooling} is reached.  In the habitable zone this steady state can produce a super-cooling of the core when tidal heating is strong enough to heat the mantle and decrease its viscosity and low enough to not dominate the surface heat flow.  This rapid cooling leads to complete core solidification, prohibiting dynamo action for most models in the habitable zone with $e>0.05$ by 10 Gyr.  In addition to weak magnetic fields, massive melt eruption rates in the habitable zone may render these fixed orbit planets {  uninhabitable}.

Commonly the term ``habitability'' refers to the influx of radiation necessary to maintain surface liquid water.  However, the full habitability of a planet must involve the dynamics of the interior and its interaction with the surface environment.
We find that tidal heating of a planetary mantle can influence surface habitability in several important ways: 

\begin{enumerate}
\item {  Prolonged magma ocean stage.}  Close-in planets with a high eccentricity ($e\gtrsim 0.1$) will experience extreme tidal heating rates of $\sim1000$ TW and tidal steady state mantle temperatures of $\sim2000$ K, implying mostly molten mantles.  These super-tidal planets are {  uninhabitable} as the surface itself is likely molten or close to the silicate solidus. \\
\item {  Extreme volcanic eruption rates.}  {  Tidal heating, in addition to increasing surface heat flow, can produce extreme surface melt production rates}.  Even if only a fraction ($\sim20\%$) of this melt erupts to the surface it can easily produce a 100 fold increase over the present-day mid-ocean ridge eruption rate ($\sim10^{13}~\mbox{kg~yr}^{-1}$).  These extreme eruption rates can lead to rapid global resurfacing and degassing that render the surface environment a violent and potentially toxic place for life.  
{  Volcanically dominated atmospheres could be significantly different from the modern-day Earth's and are potentially detectable with future space- and ground-based telescopes \citep{misra2015}.}
\\
\item {  Lack of magnetic field.}  Planetary magnetic fields are often invoked as shields necessary to maintain life.  Magnetic fields can protect the atmosphere from stellar wind erosion \citep{driscoll2013} and the surface from harmful radiation \citep{dartnell2011,griessmeier2005}.  Super-cooling of the core, which can solidify the entire core and kill the dynamo, occurs in the habitable zone after $\sim9$ Gyr with a fixed orbit.  {  Alternatively, a tidally heated stagnant-lid planet can maintain hotter mantle temperatures and lower core cooling rates, weakening the core generated magnetic field.}
Even before losing the dynamo entirely these planets may have magnetic fields that are too weak to hold the stellar wind above the atmosphere or surface.  In either case, the lack of a strong magnetic shield will be detrimental to life.
\\
\item {  Tidally driven runaway greenhouse.} {  In \S \ref{section_runaway} we show that driving a runaway greenhouse by tidal heating in the rocky interior alone is difficult.  To achieve the runaway threshold heat flux at the surface either the planet would have to be forced into a highly eccentric orbit after the mantle has cooled down to $\sim1800$ K or the dissipative material properties would have to be different.  For example, if the mantle were composed of a lower viscosity material then the maximum Maxwell tidal power could increase to $10^5$ TW (Figure \ref{contour_mu_nu}a).  A significant amount of tidal energy can also be dissipated in the liquid portions of the planet \citep{tyler2014}, which is beyond the scope of this study.
}
\end{enumerate}

With growing interest in the habitability of Earth-like exoplanets, the development of geophysical evolution models will be necessary to predict whether these planets have all the components that are conducive for life.  This paper focused on a single planet mass, but the mathematical equations can be developed to model the evolution of other rocky planet/star mass ratios, including large rocky satellites around giant planets.  
However, significant uncertainties make the application to super-Earths particularly challenging.  
The fundamental physical mechanisms underpinning plate tectonics, both in terms of its generation and maintenance over time, are not fully understood, making extrapolation to larger planets questionable.
 Perhaps most importantly,  material properties, such as viscosity, melting point, solubility, and conductivity, are poorly constrained at pressures and temperatures more extreme than Earth's lower mantle and core.  This uncertainty prevails in our own Solar System where the divergent evolution of Earth and Venus from similar initial conditions to dramatically different present-day states remains elusive.  

Future thermal-orbital modeling improvements should include coupling the evolution of the interior to the surface through volatile cycling and atmosphere stability.  Advancing the orbital model to include gravitational interactions with additional planetary companions would allow for tidal resonances, variable rotation rates, and other time-dependent orbital forcings.  In addition to the eccentricity tide explored here, an obliquity tide could also be important.
Further improvements could include dissipation in oceans or internal liquid layers, variable internal compositions, structures, and radiogenic heating rates, {  core light element depression, continental crust formation,} and eventually a direct coupling of first-principles numerical simulations.

%%%%%%%%%%%%%%%%%%%%%%%%%%%%%%%%%%%%%%%%%%%%%%%%%%%%%%%%
%%%%%%%%%%%%%%%%%%%%%%%%%%%%%%%%%%%%%%%%%%%%%%%%%%%%%%%%
%%%%%%%%%%%%%%%%%%%%%%%%%%%%%%%%%%%%%%%%%%%%%%%%%%%%%%%%
%%%%%%%%%%%%%%%%%%%%%%%%%%%%%%%%%%%%%%%%%%%%%%%%%%%%%%%%
%%%%%%%%%%%%%%%%%%%%%%%%%%%%%%%%%%%%%%%%%%%%%%%%%%%%%%%%
%%%%%%%%%%%%%%%%%%%%%%%%%%%%%%%%%%%%%%%%%%%%%%%%%%%%%%%%
%%%%%%%%%%%%%%%%%%%%%%%%%%%%%%%%%%%%%%%%%%%%%%%%%%%%%%%%
%%%%%%%%%%%%%%%%%%%%%%%%%%%%%%%%%%%%%%%%%%%%%%%%%%%%%%%%

\section*{Acknowledgements}
The authors thank W. Henning and T. Hurford for helpful discussions.  This work was supported by the NASA Astrobiology InstituteÕs Virtual Planet Laboratory under Cooperative Agreement solicitation NNH05ZDA001C.

\section*{Author Disclosure Statement}
No competing financial interests exist.

%%%%  FIGURE CAPTIONS  %%%
%%%%%%%%%%%%%%%%%%%%%%%%%%%%%%%
%%%%%%  FIGURES %%%%%%%%%%%%%%%%%

%%%  FIGURE: Tidal-Habitable Zone %%%
%%%  FIGURE: Tidal-Habitable Zone %%%

%%%  FIGURE: Heat flow contour of smod vs visc %%%

%%%  FIGURE: Heat flow vs T_mantle %%%

%%%  FIGURE: Three Models:  Q vs T_m %%%

%%%  FIGURE: Three Models: time evo %%%
\def \xone{-432} \def \xtwo{-210}
\def \yone{480} \def \ytwo{320} \def \ythree{165}
\begin{figure} \begin{center}
\includegraphics[width=0.9\textwidth]{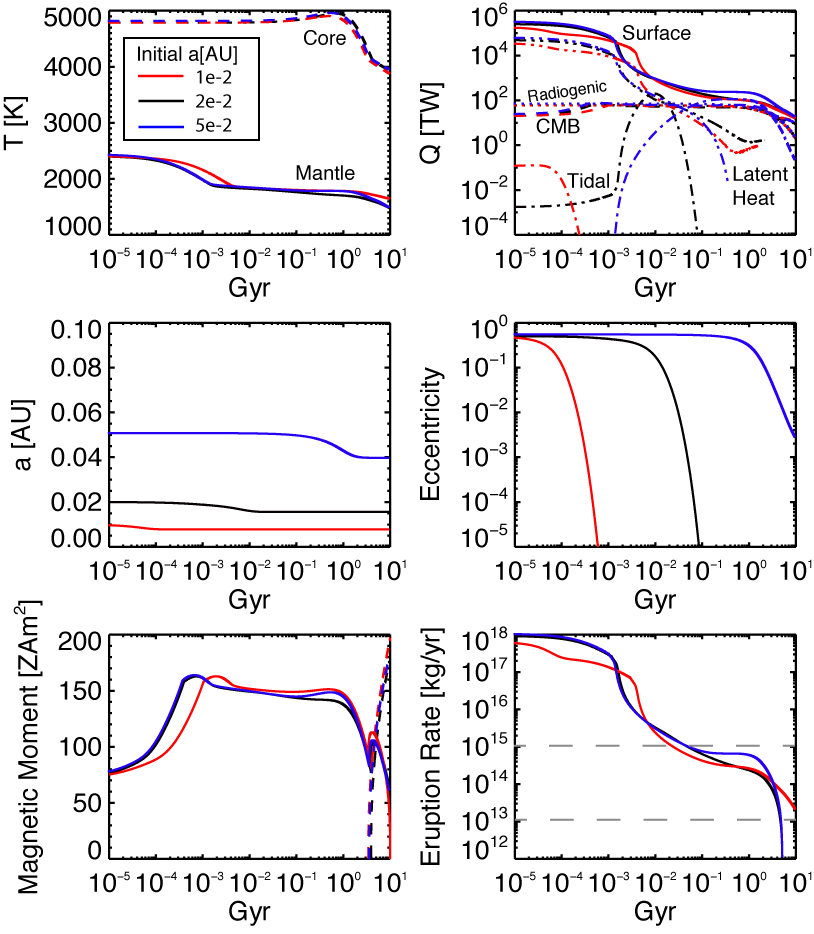}
\put(\xone,\yone){(a)}
\put(\xtwo,\yone){(b)}
\put(\xone,\ytwo){(c)}
\put(\xtwo,\ytwo){(d)}
\put(\xone,\ythree){(e)}
\put(\xtwo,\ythree){(f)}
\caption{Time evolution of models with initial orbits of $e=0.5$ and $a=0.01$ (red), $0.02$ (black), and $0.05$ (blue).  (a) Temperature in the mantle (solid) and core (dashed).  (b) Heat flow at the surface $Q_{surf}$ (solid), tidal dissipation $Q_{tidal}$ (dash-dot), mantle radiogenic heating (dotted), mantle latent heat (dash-dot-dot), and core heat flow $Q_{cmb}$ (dashed).  The runaway greenhouse heat flow threshold ($1.53\times10^{15}$ TW) is label as a solid grey line. 
(c) Orbital distance.  (d) Eccentricity.  (e) Magnetic moment of core dynamo (solid) and inner core radius (dashed).  Inner core radius axis has been scaled by core radius so the top corresponds to a completely solid core.  For reference, Earth's present day magnetic moment is about $80$ ZAm$^2$.  (f) Melt mass flux to the surface.  Melt eruption fluxes for present-day mid-ocean ridges ($10^{13}~\mbox{kg~yr}^{-1}$) and the Siberian traps ($10^{15}~\mbox{kg~yr}^{-1}$) shown for reference (grey dashed). }
\label{time_evo}
\end{center} \end{figure}

%%%  FIGURE: Contour: orbit evo %%%

%%%  FIGURE: Contour: Tidal HF %%%

%%%  FIGURE: Contour: HF surf %%%

%%%  FIGURE: Contour: HF cmb %%%

%%%  FIGURE: Contour: T man %%%

%%%  FIGURE: Contour: T core %%%

%%%  FIGURE: Contour: Magnetic moment %%%

%%%  FIGURE: Contour: melt mass flux %%%

%%%  FIGURE: Fixed orbit example %%%
\def \xone{-432} \def \xtwo{-210}
\def \yone{480} \def \ytwo{320} \def \ythree{165}
\begin{figure} \begin{center}
\includegraphics[width=0.9\textwidth]{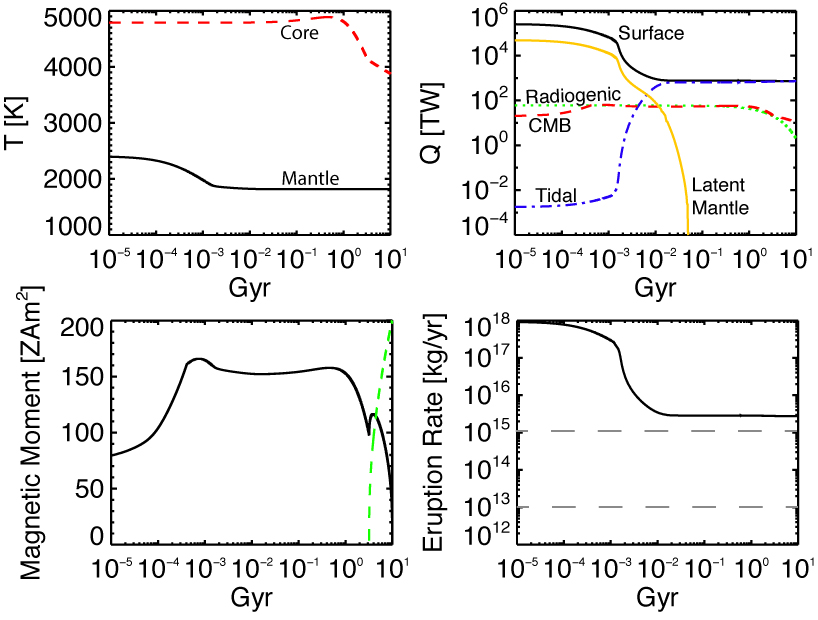}
\put(\xone,\ytwo){(a)}
\put(\xtwo,\ytwo){(b)}
\put(\xone,\ythree){(c)}
\put(\xtwo,\ythree){(d)}
\caption{Time evolution of a model with fixed orbit of $e=0.5$ and $a=0.02$ AU.  (a) Temperature in the mantle (solid) and core (dashed).  (b) Heat flow at the top of the mantle $Q_{conv}$ (solid), tidal $Q_{tidal}$ (dash-dot), mantle radiogenic heating (dotted), and core heat flow $Q_{cmb}$ (dashed).  (c) Magnetic moment of core dynamo (solid) and inner core radius (dashed).  Inner core radius axis goes from zero to total core radius.  For reference, Earth's present day magnetic moment is about $80$ ZAm$^2$.  (d) Melt mass flux to the surface.  Melt eruption fluxes for present-day mid-ocean ridges ($10^{13}~\mbox{kg~yr}^{-1}$) and the Siberian traps ($10^{15}~\mbox{kg~yr}^{-1}$) shown for reference (grey dashed).}
\label{fixed_example}
\end{center} \end{figure}
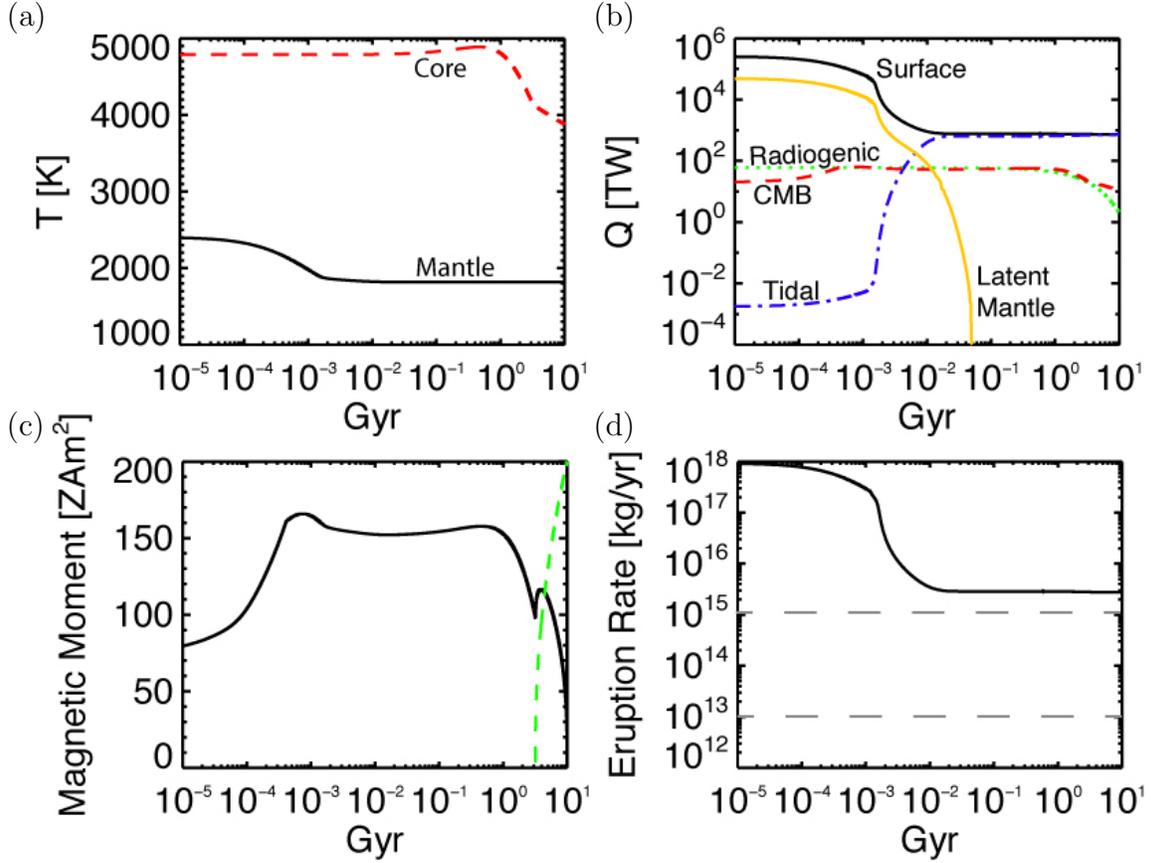
%%%%%%%%%%%%%%%%%%%%%%%%%%%%%%%%%%%%%

%%%  FIGURE: Contour: t in tidal runaway %%%

%%%  FIGURE: Contour: time (a) of tidal dominated, (b) tidal-runaway %%%

\pagebreak\pagebreak\pagebreak

%%%%%%%%%%%%%%%%%%%%%%%%%%%%
%%%%%  APPENDIX %%%%%%%%%%%%%%%%
\section*{Appendix}
\appendix
\renewcommand\thefigure{\thesection.\arabic{figure}}    
\setcounter{figure}{0}    

%%%%%%%%%%%%%%%%
\section{Tidal Dissipation Model}  \label{appendix_dissipation}
\begin{figure} \begin{center}
\includegraphics[width=0.8\linewidth]{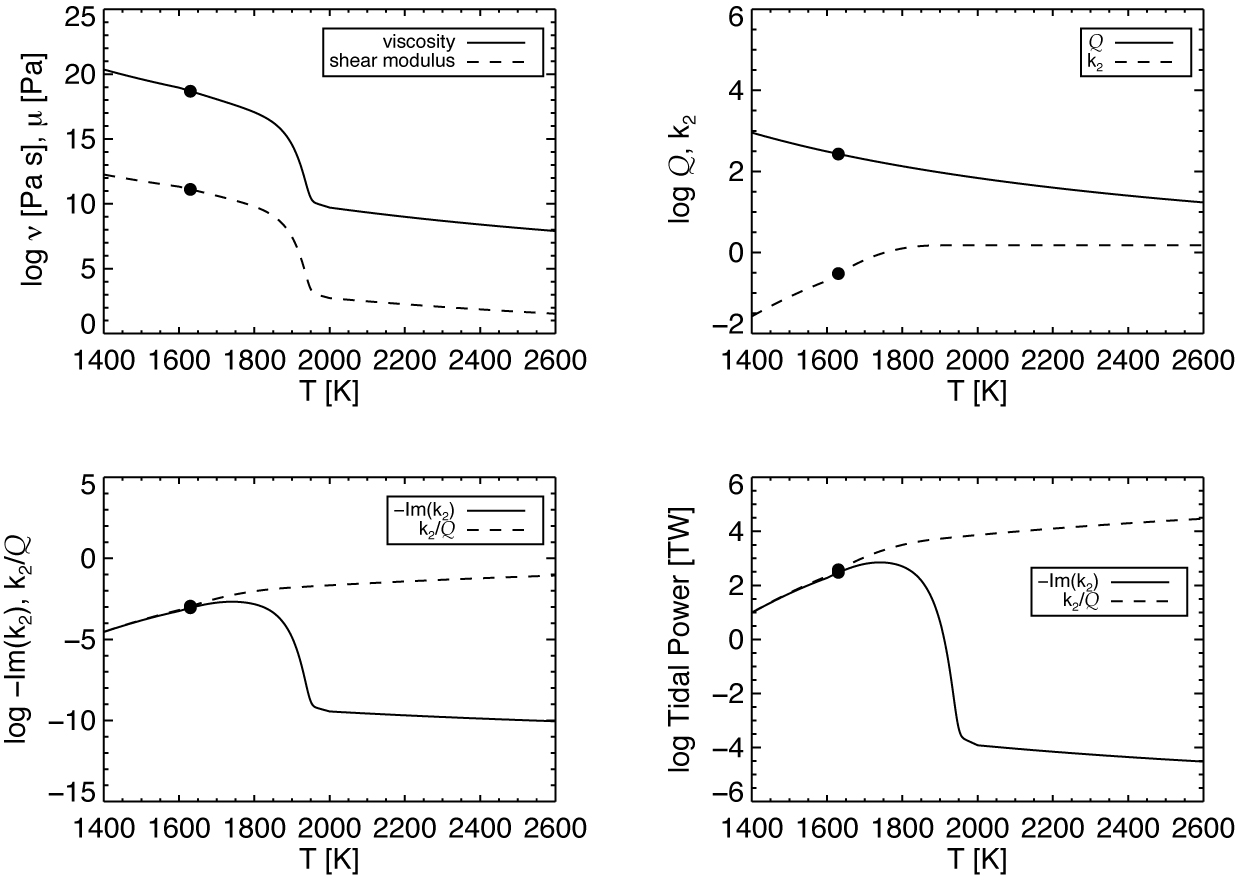}
\put(-380,265){(a)}
\put(-180,265){(b)}
\put(-380,125){(c)}
\put(-180,125){(d)}
\caption{Tidal dissipation parameters as a function of mantle temperature $T_m$ for the nominal shear modulus activation energy of $A_\mu=2\times10^5~\jmol$.  Circles denote the chosen calibration points for the present-day mantle.}
\label{tidal_t}
\end{center} \end{figure}

This section demonstrates the dependence of the tidal dissipation model and material properties on mantle temperature.
Figure \ref{tidal_t} shows several parameters related to the tidal dissipation rate as a function of mantle temperature for the nominal shear modulus activation energy of $A_\mu=2\times10^5~\jmol$.  The Maxwell model that uses the full form of $-Im(k_2)$ in (\ref{im_k2}) differs from the common approximation of $-Im(k_2)\approx k_2/\mathcal{Q}$ for mantles hotter than the present-day ($T_m>1630$ K) (Figure \ref{tidal_t}c).  The difference between the Maxwell model and this approximation corresponds to about 10 orders of magnitude larger tidal heat flow at high temperature (Figure \ref{tidal_t}d).  The approximation is invalid at high temperature because it does not account for the drop in tidal dissipation expected in a liquid, since the approximation relies on $\mathcal{Q}\propto \eta/\mu$, which is constant, whereas the Maxwell model predicts a sharp drop in tidal dissipation with viscosity when $\mu/\beta <<2/10$.

%%% FIG

%%%%%%%%  SECTION: THERMAL HISTORY MODEL  %%%%%%%%%
\section{Thermal History Model}   \label{appendix_thermalmodel}
\subsection{Geotherm}
The mantle temperature profile is assumed to be adiabatic everywhere except in the thermal boundary layers where it is conductive.  The adiabatic temperature profile in the well mixed region of the mantle is approximated to be linear in radius, which is a good approximation considering that mantle thickness $D=2891$ km is much less than the adiabatic scale height $H=c_p/\alpha g\approx12650$ km,
\be 
T_{ad}=T_{UM} + \gamma_{ad}(R-r-\delta_{UM})~,
\label{T_ad} \ee
where the adiabatic gradient is $\gamma_{ad}\approx0.5~\mbox{K/km}$.  In the thermal boundary layers the conductive temperature solutions,
\begin{eqnarray}
\Delta T_{UM} \mbox{erf}\left[ \frac{R-r}{\delta_{UM}}\right] + T_s ~~,~~ \mbox{Upper mantle} \\
\Delta T_{LM} \mbox{erf}\left[ \frac{R_c-r}{\delta_{LM}}\right] + T_{cmb} ~~,~~ \mbox{Lower mantle}
\label{T_erf} \end{eqnarray}
replace the adiabat.  Thermal boundary layer temperature jumps are $\Delta T_{UM}=T_{UM}-T_g$ and $\Delta T_{LM}=T_{cmb}-T_{LM}$, and thermal boundary layer depth is $\delta$.
Figure \ref{figure_geotherm} shows an example whole planet geotherm $T(r)$ at four times in the evolution.
Surface temperature $T_g$ is assumed to be equal to the equilibrium temperature,
\be T_{eq}=\left( \frac{L_*}{16\pi\sigma a^2}\right)^{1/4} ~, \label{t_eff} \ee
where $L_*$ is stellar luminosity and $\sigma$ is the Stefan-Boltzmann constant.

%%% FIG

The core temperature profile is assumed to be adiabatic throughout the entire core, i.e.\ the thermal boundary layers within the core are ignored.  This is a good approximation because the low viscosity and high thermal conductivity of liquid iron produce very small thermal boundary layers that are insignificant on the scale of the whole planet.  The core adiabatic profile is approximated by 
\begin{equation}
T_c(r)=T_{cmb}\exp \left( \frac{R_c^2-r^2}{D_N^2} \right)~,
\label{T_c} \end{equation}
where $D_N\approx 6340$ km is an adiabatic length scale \citep{labrosse2001}.  The iron solidus is approximated by Lindemann's Law,
\begin{equation}
T_{Fe}=T_{Fe,0} \exp \left[  -2 \left( 1- \frac{1}{3\gamma_c} \right) \frac{r^2}{D_{Fe}^2} \right]~,
\label{lindemann} \end{equation}
where $T_{Fe,0}=5600$ K, $\gamma_c$ is the core Gruneisen parameter, and $D_{Fe}=7000$ km is a constant length scale \citep{labrosse2001}.  
This simple treatment of the core solidus does not account for volatile depression of the solidus, which has been demonstrated experimentally \citep{hirose2013}, and would act to slow inner core growth.  Inner core radius can then be solved for by finding the intersection of (\ref{T_c}) and (\ref{lindemann}).  For details see \citet{driscoll2014}.

\subsection{Mantle and Core Heat Flows}
In this section we define the remaining heat flows that appear in the mantle (\ref{mantle_energy}) and core (\ref{core_energy}) energy balance.  

The convective cooling of the mantle $Q_{conv}$ is proportional to the temperature gradient in the upper mantle thermal boundary layer,
\begin{equation}
Q_{conv}=A k_{UM} \frac{\Delta T_{UM}}{\delta_{UM}}~,
\label{q_conv}\end{equation}
where $A$ is surface area and $k_{UM}$ is upper mantle thermal conductivity.  $Q_{conv}$ is written in terms of $T_m$ and the thermal boundary layer thickness $\delta_{UM}$ by requiring that the Rayleigh number of the boundary layer $Ra_{UM}$ be equal to the critical Rayleigh number for thermal convection $Ra_c\approx660$ \citep{howard1966,solomatov1995,sotin1999,driscoll2014}. This constraint gives,
\begin{equation}
Q_{conv}=A k_{UM}  \left( \frac{\alpha g}{Ra_c \kappa }\right)^{\beta} (\epsilon_{UM}\Delta T_m)^{\beta+1}  (\nu_{UM})^{-\beta}~,
\label{q_conv_m} \end{equation}
where the thermal boundary layer temperature jump $\Delta T_{UM}$ has been replaced by $\Delta T_{UM}\approx\epsilon_{UM}\Delta T_m$, 
$\epsilon_{UM}=\exp(-(R_{UM}-R_m)\alpha g/c_p)\approx0.7$ is the adiabatic temperature decrease from the average mantle temperature to the bottom of the upper mantle thermal boundary layer, $\Delta T_m=T_m-T_g$, and the mantle cooling exponent is $\beta=1/3$.

Radiogenic heat production in the Earth is generated primarily by the decay of $^{238}U$, $^{235}U$, $^{232}Th$, and $^{40}K$, which is approximated in the mantle by,
\begin{equation}
Q_{rad}(t)=Q_{rad,0} \exp(-t/\tau_{rad})~,
\label{q_rad} \end{equation}
where $Q_{rad,0}$ is the initial radiogenic heat production rate at $t=0$ and $\tau_{rad}$ is the radioactive decay time scale that approximates the decay of the four major isotopes.  
The precise bulk silicate Earth radiogenic heat production rate is somewhat uncertain, so we use a nominal value of $Q_{rad}(t=4.5~\mbox{Gyr})=13$ TW \citep{jaupart2007}.

Similar to the mantle convective heat flow, the CMB heat flow is,
\begin{equation}
Q_{cmb}=A_c k_{LM} \frac{\Delta T_{LM}}{\delta_{LM}}  ~,
\label{q_cmb} \end{equation}
where $A_c$ is core surface area and $k_{LM}$ is lower mantle thermal conductivity. The lower mantle and CMB temperatures, $T_{LM}$ and $T_{cmb}$, are extrapolations along the mantle and core adiabats: $T_{LM}=\epsilon_{LM}T_m$ and $T_{cmb}=\epsilon_c T_c$, where  $\epsilon_{LM}=\exp(-(R_{LM}-R_{m})\alpha g /c_p) \approx 1.3$ 
and $\epsilon_c\approx0.8$. 
The lower mantle thermal boundary layer thickness is also derived by assuming the boundary layer Rayleigh number is critical and that  $\nu_{LM}=2\nu_{UM}$, which was found by \citet{driscoll2014} to produce a nominal Earth model.

Core secular cooling is
\begin{equation}
Q_{core}=-M_c c_c \dot{T}_c~,
\label{q_core} \end{equation}
where $M_c$ is core mass, $c_c$ is core specific heat, and $\dot{T}_c$ is the rate of change of the average core temperature $T_c$.  

Radiogenic heat in the core is produced primarily by the decay of $^{40}K$ \citep{gessmann2002,murthy2003,corgne2007}.  
Its time dependence is treated the same as mantle radiogenic heat in (\ref{q_rad}), but with a radioactive decay time scale of $\tau_{rad,c}=1.2$ Gyr.  We assume an abundance of $^{40}K$ in the core that corresponds to 2 TW of heat production after 4.5 Gyr.

\subsection{Melting}  \label{appendix_solidus}
The mantle solidus is approximated by a third-order polynomial \citep{elkinstanton2008b},
\be
T_{sol}(r)=A_{sol}r^3+B_{sol}r^2+C_{sol}r+D_{sol}~,
\label{solidus} \ee
where the coefficients are constants (see Table \ref{table1}).  This solidus is calibrated to fit the following constraints: solidus temperature of $1450$ K at the surface, solidus temperature of $4150$ K at the CMB \citep{andrault2011}, and present-day upwelling melt fraction of $f_{melt}=8\%$.  The liquidus is assumed to be hotter by a constant offset $\Delta T_{liq}=500$ K, so $T_{liq}(r)=T_{sol}(r)+\Delta T_{liq}$.

Mantle melt heat loss (or advective heat flow) is modeled as,
\begin{equation}
Q_{melt}=\epsilon_{erupt}\dot{M}_{melt} \left( L_{melt} + c_{m} \Delta T_{melt} \right)~,
\label{q_melt} \end{equation}
where $\epsilon_{erupt}=0.2$ is the efficiency of magma eruption to the surface (assumed to be constant and equal to present-day value), $\dot{M}_{melt}$ is melt mass flux (see below),
$L_{melt}$ is latent heat of the melt, $c_{m}$ is specific heat of the melt, and $\Delta T_{melt}$ is the excess temperature of the melt at the surface (see below).  This formulation of heat loss is similar to the "heat pipe" mechanism invoked for Io \citep{oreilly1981,moore2003}, where melt is a significant source of heat loss.   We note that this mechanism is more important for stagnant lid planets where the normal conductive heat flow is lower \citep{driscoll2014}.

The melt mass flux $\dot{M}_{melt}$ is the product of the upwelling solid mass flux times the melt mass fraction $f_{melt}$,
\begin{equation}
\dot{M}_{melt}=\dot{V}_{up} \rho_{solid} f_{melt}(z_{UM})~,
\label{m_melt}\end{equation}
where solid density is $\rho_{solid}$, volumetric upwelling rate is $\dot{V}_{up}=1.16\kappa A_p/ \delta_{UM}$, $z_{UM}=R-\delta_{UM}$, and melt fraction is
\be f_{melt}(z)=\frac{T_m(z)-T_{sol}}{T_{liq}-T_{sol}} ~. \label{f_melt} \ee  
This model predicts a ridge melt production of $\dot{M}_{melt}=2.4\times10^6~\mbox{kg~s}^{-1}$ for $\delta_{UM}=80$ km and $f_{melt}=0.1$, similar to present-day global melt production estimates \citep{cogne2004}.

\begin{figure} \begin{center}
\includegraphics[width=0.8\linewidth]{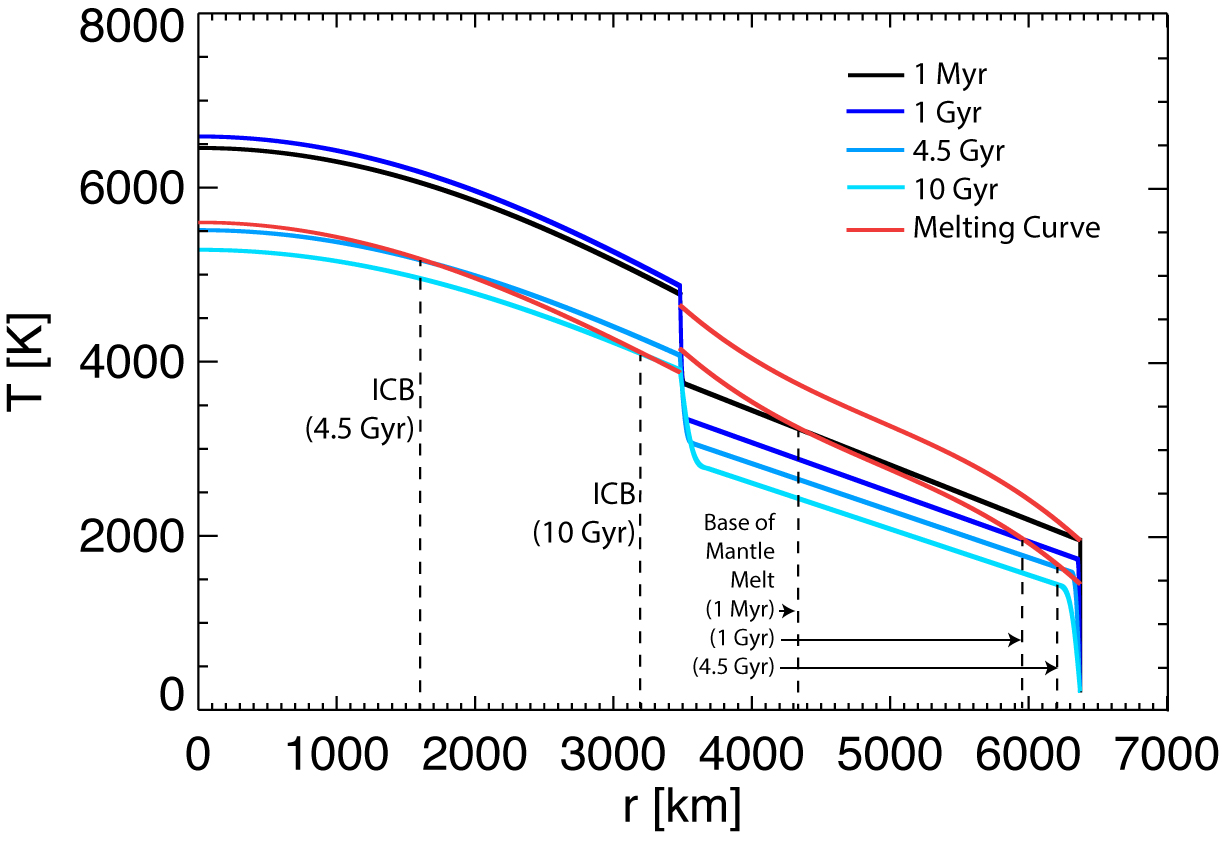}
\caption{Geotherm temperature profile $T(r)$ for the orbit evolving model with initial orbit of $a=0.05$ AU and $e=0.5$.  Temperature profiles are shown at 1 Myr, 1 Gyr, 4.5 Gyr, and 10 Gyr.  The core liquidus and mantle liquidus and solidus are shown in red.}
\label{figure_geotherm}
\end{center} \end{figure}

We define the magma ocean as the region of the mantle with temperature exceeding the liquidus.  Given the geotherm in (\ref{T_ad},\ref{T_erf}) and the liquidus $T_{liq}(r)$ similar to (\ref{solidus}), the mantle will mainly freeze from the bottom of the convecting mantle up because the liquidus gradient is steeper than the adiabat \citep[e.g.][]{elkinstanton2012}.  However, if the core is hot enough a second melt region exists in the lower mantle boundary layer, where the temperature gradient exceeds the liquidus and the mantle freezes towards the CMB.  As can be seen in Figure \ref{figure_geotherm}, a basal magma ocean exists for about 4 Gyr before solidifying.

Latent heat released from the solidification of the mantle is
\be
Q_{L,man}=\dot{M}_{sol} L_{melt}~,
\label{Q_latent_man} \ee
where $L_{melt}$ is the latent heat released per kg and $\dot{M}_{sol}$ is the solid mantle growth rate.  The growth rate is calculated assuming a uniform mantle density $\rho_m$ so that $\dot{M}_{sol}=\rho_m \dot{V}_{sol}$, where $\dot{V}_{sol}=-\dot{V}_{liq}$.  The rate of change of the liquid volume of the mantle is 
\be \dot{V}_{liq} =\frac{dV_{liq}}{dT_m} \dot{T}_m~,
\label{dot_v_liq} \ee
where $\dot{T}_m$ is the mantle secular cooling rate and $dV_{liq}/dT_m$ is linearly approximated by $8\times10^{17}~\mbox{m}^3\mbox{K}^{-1}$, which is the change in liquid volume from a $90\%$ liquid to a completely solid mantle.
This approximation implies that the latent heat released due to mantle solidification is linearly proportional to the mantle secular cooling rate, and the ratio of the latent heat flow to the mantle secular cooling heat flow is $Q_{L,man}/Q_{sec,m}\approx0.24$.  For example, a mantle solidification time of 100 Myr corresponds to an average latent heat release of $Q_{L,man}\approx400$ TW over that time.

\subsection{Core Dynamo}
Given the thermal cooling rate of the core, the magnetic dipole moment $\mathcal{M}$ is estimated from the empirical scaling law,
\be \mathcal{M} = 4\pi R_c^3 \gamma_d \sqrt{\rho/2\mu_0} \left( F_c D_c \right)^{1/3} \label{magmom} \ee
where $\gamma_d=0.2$ is the saturation constant for fast rotating dipolar dynamos, $\mu_0=4\pi \times 10^{-7} \mbox{H~m}^{-1}$ is magnetic permeability, $D_c=R_c-R_{ic}$ is the dynamo region shell thickness, $R_c$ and $R_{ic}$ are outer and inner core radii, respectively, and $F_c$ is the core buoyancy flux \citep{olson2006}.  
We assume that the field is dipolar, ignoring the complicating influences of shell thickness and heterogeneous boundary conditions \citep[e.g.][]{heimpel2005,driscoll2009b,aubert2009,olson2014}.  In this formulation a positive buoyancy flux implies dynamo action, which is a reasonable approximation when the net buoyancy flux is large, but may overestimate the field strength at low flux.
The total core buoyancy flux $F_c$ is the sum of thermal and compositional buoyancy fluxes,
\be F_c=F_{th}+F_\chi \label{f_c}\ee
where the thermal and compositional buoyancy fluxes are
\begin{eqnarray} F_{th}&=&\frac{\alpha_c g_c}{\rho_c c_c} q_{c,conv} \label{f_th}  \\
	F_\chi &=& g_i\frac{\Delta \rho_\chi}{\rho_c} \left(\frac{R_{ic}}{R_c}\right)^2 \dot{R}_{ic} ~, \label{f_chi}
\end{eqnarray}
where the subscript $c$ refers to bulk core properties, core convective heat flux is $q_{c,conv}=q_{cmb}-q_{c,ad}$, gravity at the ICB is approximated by $g_{ic}=g_c R_{ic}/R_c$, and the outer core compositional density difference is $\Delta \rho_\chi=\rho_c-\rho_\chi$ with $\rho_\chi$ the light element density.  For simplicity, the expression for light element buoyancy (\ref{f_chi}) ignores buoyancy due to latent heat release at the ICB because it is a factor of $3.8$ less than buoyancy of the light elements.

 The isentropic core heat flux at the CMB, proportional to the gradient of (\ref{T_c}), is
\be q_{c,ad}=k_c T_{cmb} R_c/D_N^2 ~,\label{q_cad} \ee
where core thermal conductivity is approximated by the Wiedemann-Franz law,
\be
 k_c=\sigma_c L_c T_{cmb} ~,\label{core_conductivity}
 \ee 
 and electrical conductivity is $\sigma_c$ and $L_c$ is the Lorentz number.  For typical values of high pressure-temperature iron, $\sigma_c=10\times10^5~\ohmm$ \citep{pozzo2012,gomi2013}, $L_c=2.5\times10^{-8}~\mbox{W}\Omega\mbox{K}^{-1}$, and $T_{cmb}=4000$ K, the core thermal conductivity is $k_c=100~\mbox{Wm}^{-1}\mbox{K}^{-1}$.

%%%%%%%  APPENDIX  FIGURES  %%%%%%
%%%%%  FIGURE

%%%  FIGURE: Geotherm %%%
%%%%%%%%%%%%%%%%

%%%%  Use Dropbox file %%%
\bibliographystyle{/Users/peterdriscoll/Dropbox/latex/elsart-harv2}
\bibliography{/Users/peterdriscoll/Dropbox/latex/bibtex}

%%%%%  TABLE OF CONSTANTS.   %%%%%%%
%%%%  TABLE 1:  Model constants    %%%%%
%%%% TABLE of CONSTANTS %%%%

\begin{longtable}{l | l | l | l}
Symbol & Value & Units & Reference \\
\hline
\endhead
   $A_\nu$      &           $3\times10^5$              &  J mol$^{-1}$ &	 Viscosity activation energy in (\ref{nu})	\\
   $A_\mu$	&	$2\times10^5$		&	J mol$^{-1}$ &	 Nominal shear modulus activation energy in (\ref{mu})	\\
$A_{sol}$		&    $-1.160\times10^{-16}$	& K$/$m$^{3}$	&	Solidus coefficient in (\ref{solidus}) (ET08) \\
      $\alpha$ &       $3\times10^{-5}$    & K$^{-1}$ &	  Thermal expansivity of mantle	\\
   $\alpha_c$ &       $1\times10^{-5}$    & K$^{-1}$ &	  Thermal expansivity of core	\\
	$B$		&	$2.5$			& nd		& 	Melt fraction coefficient in (\ref{epsilon_phase}) \\
   $B_{sol}$		&    $1.708\times10^{-9}$	& K$/$m$^{2}$	&	Solidus coefficient in (\ref{solidus}), calibrated \\
   $\beta$ 	&	$1/3$			& nd		&	Convective cooling exponent in (\ref{q_conv_m})\\
   $\beta_{st}$	&	$1.71\times10^{4}$	&	GPa		& Effective mantle stiffness, calibrated in \S \ref{section_tidal_model} \\
      $c_m$          &           1265    & J kg$^{-1}$ K$^{-1}$ &	  Specific heat of mantle	\\
   $c_c$         &           840    & J kg$^{-1}$ K$^{-1}$ &	  Specific heat of core	\\
      $C_{sol}$		&    $-9.074\times10^{-3}$	& K$/$m	&	Solidus coefficient in (\ref{solidus}), calibrated \\
      $D$           &       2891    & km &	  Mantle depth	\\
     $D_{Fe}$              &       7000     & km &	  Iron solidus length scale	\\
  $D_N$              &       6340    & km &	  Core adiabatic length scale	\\
     $D_{sol}$		&    $1.993\times10^{4}$	& K	&	Solidus coefficient in (\ref{solidus}), calibrated \\
     $\delta_{ph}$		&	$6$	& nd &	Rheology phase coefficient in (\ref{epsilon_phase}, \ref{epsilon_phase2}) \\
         $E_G$     &           $3\times10^5$    &  J kg$^{-1}$ &	  Gravitational energy density release at ICB 	\\
%   $\eta_{UM}$        &          0.7   & nd &	  Upper mantle adiabatic temperature drop	\\
%   $\eta_{LM}$        &           1.3    & nd &	 Lower mantle adiabatic temperature jump	\\
%   $\eta_c$        &          0.8    & nd &	  Average core to CMB adiabatic temperature drop	\\
   $\epsilon_{UM}$        &          0.7   & nd &	  Upper mantle adiabatic temperature drop	\\
   $\epsilon_{LM}$        &           1.3    & nd &	 Lower mantle adiabatic temperature jump	\\
   $\epsilon_c$        &          0.8    & nd &	  Average core to CMB adiabatic temperature drop	\\
     $\phi^*$		&	$0.8$	& nd &	Rheology phase coefficient in (\ref{epsilon_phase}, \ref{epsilon_phase2}) \\
   $g_{UM}$          &           $9.8$  & $\mbox{m~s}^{-2}$ 	&	Upper mantle gravity	\\
   $g_{LM}$          &           $10.5$ & $\mbox{m~s}^{-2}$ 	 &	  Lower mantle gravity	\\
   $g_c$          &           $10.5$     & $\mbox{m~s}^{-2}$ 	 &	  CMB gravity	\\
     $\gamma_{c}$      &        $1.3$            & nd  &	  Core Gruneisen parameter	\\   
      $\gamma_{dip}$       &          0.2    & nd &	  Magnetic dipole intensity coefficient in (\ref{magmom})	\\
     $\gamma_{ph}$		&	$6$	&	 nd &  Rheology phase coefficient in (\ref{epsilon_phase}, \ref{epsilon_phase2}) \\
   $k_{UM}$          &           $4.2$    & W m$^{-1}$ K$^{-1}$ &	  Upper mantle thermal conductivity	\\
   $k_{LM}$         &           $10$    &  W m$^{-1}$ K$^{-1}$ &	  Lower mantle thermal conductivity	\\
   $\kappa$	&		$10^6$	&	m$^2$ s$^{-1}$	&	Mantle thermal diffusivity \\
      $L_{Fe}$          &           750    & kJ kg$^{-1}$ &	  Latent heat of inner core crystallization	\\
         $L_{melt}$     &           320    & kJ kg$^{-1}$  &	  Latent heat of mantle melting	\\   
            $L_e$         &       $2.5\times10^{-8}$                            &  W $\Omega$ K$^{-1}$&	  Lorentz number	\\             
        $L_*$	&	$3.09\times10^{23}$		&	W		& Stellar luminosity for $M_*=0.1M_{sun}$ (B13)  \\
   $M_m$           &       $4.06\times10^{24}$    &   kg  &	  Mantle mass	\\
   	$M_c$	&	$1.95\times10^{24}$		& kg		&	Core mass \\
	$\mu_{ref}$	&	$6.24\times10^4$	&	Pa	&	Reference shear modulus in (\ref{mu}) \\
      $\mu_0$            &       $4\pi\times10^{-7}$     & H m$^{-1}$ &	  Magnetic permeability	\\
%      $\nu_0$        &       $6\times10^7$    &  m$^2$s$^{-1}$ &	  Reference viscosity	\\
      $\nu_{ref}$        &       $6\times10^7$    &  m$^2$s$^{-1}$ &	  Reference viscosity	\\
   $\nu_{LM}/\nu_{UM}$   &    $2$          & nd &	  Viscosity jump from upper to lower mantle	\\
%   $\nu_{m}/\nu_{UM}$ &           10    &  nd &	    Viscosity jump from upper to mid-mantle	\\
   $Q_{rad,0}$           &       $60$    & TW &	 Initial mantle radiogenic heat flow (J07) \\
   $R$         &       6371    &  km &	  Surface radius 	\\
   $R_c$          &       3480    & km  &	  Core radius	\\
   $R_m$		&	4925		& km		& Radius to average mantle temperature $T_m$\\  % Note: R_mm=4925 gives eta_um=0.73.
   $Ra_c$             &           660    & nd  &	  Critical Rayleigh number	\\
   $\rho_c$        &           11900    & kg m$^{-3}$ &	  Core density	\\
   $\rho_{ic}$          &           13000    & kg m$^{-3}$ &	  Inner core density	\\
   $\rho_m$         &           4800    & kg m$^{-3}$ &	  Mantle density	\\
      $\rho_{melt}$         &           2700    & kg m$^{-3}$ &	  Mantle melt density	\\
      $\rho_{solid}$       &           3300                   & kg m$^{-3}$ &	  Mantle upwelling solid density 	\\   
   $\Delta \rho_{\chi}$   &            700    & kg m$^{-3}$  &	  Outer core compositional density difference	\\
  $\sigma_c$      &           $10\times10^5$    & S m$^{-1}$ &	  Core electrical conductivity	\\
   $T_{Fe,0}$           &           5600    & K  &	  Iron solidus coefficient in (\ref{lindemann})	\\
      $\tau_{rad}$    &       $2.94$    & Gyr &	  Mantle radioactive decay time scale	\\
            $\tau_{rad,c}$    &       $1.2$    & Gyr &	  Core radioactive decay time scale	\\
            $\xi$		&	$5\times10^{-4}$	& nd &	Rheology phase coefficient in (\ref{epsilon_phase}, \ref{epsilon_phase2}) \\
\caption{Model constants.   Non-dimensional units are denoted n.d.  References are: B13=\citet{barnes2013}; ET08=\citet{elkinstanton2008b}; H00=\citet{hirschmann2000}; J07=\citet{jaupart2007}; M84=\citet{mckenzie1984}; M88=\citet{mckenzie1988}.} 
\label{table1}
\end{longtable}

%%%%%  TABLE OF CONSTANTS.   %%%%%%%

\end{document}